\documentclass[pre,twocolumn]{revtex4-1}

\usepackage{graphicx}
\usepackage{amsmath}
\usepackage{mathtools}
\usepackage{amssymb}

\usepackage{verbatim}   
\usepackage{color}      
\usepackage{subfigure}  
\usepackage{hyperref}   

\usepackage{braket}

\newcommand{\ba}{\begin{eqnarray}}
\newcommand{\ea}{\end{eqnarray}}
\newcommand{\be}{\begin{equation}}
\newcommand{\ee}{\end{equation}}

\begin{document}

\title{Soft-particle lattice-gas in 1d:  one- and two-component cases}

\author{Derek Frydel}
\affiliation{Department of Chemistry, Federico Santa Maria Technical University, Campus San Joaquin, Santiago, Chile\\}
\author{Yan Levin}
\affiliation{Institute of Physics, The Federal University of Rio Grande do Sul, Porto Alegre 91501-970, Brazil}

\date{\today}

\begin{abstract}
The object of the present article is a 1d lattice-gas system comprised of soft-particles, wherein 
particles interact only if they occupy the same or a neighboring site, 
as a simple representation of penetrable particles of soft condensed matter.  
To represent different scenarios, two different realizations of the lattice model are considered, 
a one-component and a two-component system, where in the two-component case particles
of the same species repel and those of opposite species attract each other.  
The systems are analyzed entirely within the transfer matrix framework.  
Special attention is paid to the criterion devised in 
Ref. [{\sl Phys. Rev. E} {\bf 63}, 031206 (2001)], which serves to separate two 
classes of behavior encountered in a one-component penetrable particle systems.  
In addition to confirm the existence of a similar criterion for the one-component lattice-gas model, 
we find that the same criterion can be applied to the two-component system to provide 
conditions for thermodynamic catastrophe.  
\end{abstract}

\pacs{
}

\maketitle

\section{Introduction}

The present work investigates a 1d lattice model of soft particles.  Soft interactions imply 
the possibility of multiple occupations of a single site, and the model is intended to be a 
simple representation of penetrable particles, such as the Gaussian core or the 
Penetrable-Sphere model.

In soft condensed matter, penetrable pair potentials have been recognized as realistic 
representations of effective interactions between a number of large macromolecules. 
For example, the Penetrable-Sphere model represents micelles in a solvent \cite{Witten89}, 
while the Gaussian core model accurately captures entropic repulsions between self-avoiding 
polymer coils in a good solvent \cite{Hansen00,Hansen01}.  The the generalized exponential 
model, $\exp(-r^4)$, accurately describes the effective repulsion between flexible dendrimers 
in a solution \cite{Mladek07,Mladek08a,Mladek08b,Mladek10}. 
Other soft potentials have been 
suggested to include an ever wider class of macromolecules \cite{Likos01a}.  Penetrability
has been further extended to include charged macromolecules, represented by a 
divergence-free Coulomb potential \cite{Hansen12,Frydel13,Frydel16}.

Due to the absence of hard-core interactions, the density of penetrable systems may acquire 
arbitrarily large values.  This in turn leads to new behaviors not seen in standard models.  
For example, a solid phase 
of Gaussian particles melts in two ways:  when pressure is reduced, as in normal melting, and 
when pressure is increased, known as reentrant melting \cite{Saija05,Saija05b,Ikeda13}.

Penetrable-spheres, on the other hand, do not undergo reentrant melting.  A solid phase is 
preserved all the way into infinite densities. A lattice structure, instead of being shrunk or 
deformed upon compression in order to create additional sites, remains intact and excess 
of particles is accommodated by allowing multiple occupation of existing lattice sites.  Such ``stacks'' 
of several particles sharing the same space can form already in a dense liquid phase prior 
to freezing.   Advantage of a ``stacked'' over a more uniformly distributed system 
is that the ``stacking'' arrangement reduces a number of overlaps by minimizing interactions 
between particles in different ``stacks''.  This, in turn, lowers the system energy.

Between the Gaussian core and the Penetrable-Sphere model there exists a continuum of 
functional forms of penetrable potentials that fall into one of the two classes of behavior.  
Using the mean-field analysis, Likos {\sl et al} \cite{Likos01} determined criterion for 
predicting a type of behavior for any given pair potential.   
If the Fourier transform of the pair interaction  
is everywhere positive, then one expects the Gaussian model like behavior.  If, on the other 
hand, it is not positive everywhere, the system exhibits ``stacking'' formations as in the 
Penetrable-Sphere model.

The leading motivation for the present work is to shed light on the two classes of behavior 
and to better understand the Likos-Lang-Watzlawek-L\"owen (LLWL) criterion in the context of 
a simple lattice-gas system.  The model consists of a 1d array of discrete sites.  There 
is no bound on how many particles may occupy a single site.  Particles interact only if they 
are on the same or a neighboring site.  The interaction strength is regulated with two 
parameters, $K$ for interactions between particles on the same site, and $K'=\alpha K$, 
for interactions between particles occupying neighboring sites.

To analyze the model and its properties, we use the transfer matrix 
method.  Within this methodology a given system is characterized by a transfer matrix.  All the 
thermodynamic quantities can then be expressed in terms of the transfer matrix 
eigenvalues and eigenvectors. As the occupation number is unlimited, the transfer matrix has 
infinite size.  In practice, however, a $20\times 20$ matrix suffices for most situations.  
Eigenvalues and eigenvectors of the transfer matrix 
are then calculated numerically using any of the standard software packages, 
such as Mathematica or Matlab. 

In addition to a one-component scenario, the work considers a 
two-component system, where particles of the same species repel and those of opposite 
species attract each other.  Such systems have been considered for penetrable-spheres  
\cite{Frydel17} and the Gaussian core model  \cite{Frydel18}.  As the two-component Gaussian 
core model is well behaved, the two-component Penetrable-Sphere model is thermodynamically 
unstable.  We show that the LLWL criterion of classification  
\cite{Likos01} can be extended and applies to a two-component system, 
where it serves as a criterion of thermodynamic stability.

The work is organized as follows.  In Sec. (\ref{sec:noninteractive}) we consider a 1d 
lattice model for non-interactive particles. Different ways of counting configurations are
considered, leading to different partition functions.  In Sec. (\ref{sec:model1}) we consider 
a one-component lattice model with soft repulsive interactions.   Here we determine the 
existence of two types of behavior, in agreement with 
penetrable particles of soft condensed matter. 
In Sec. (\ref{sec:model2}) we consider a two-component system with particles of the
same species repelling and of different species attracting each other.  We show
that the LLWL criterion of a one-component case apply 
to the two-component system as a criterion of thermodynamic stability.  
Finally, in Sec. (\ref{sec:conclusion}) we conclude the work.

\section{non-interactive particles}
\label{sec:noninteractive}

Given a 1d array of $L$ lattice sites and $N$ indistinguishable and non-interactive particles, 
the canonical partition function, for the case where at most one particle can occupy a lattice site,  
is 
\be
Z = \frac{L!}{N!(L-N)!}.  
\label{eq:Z_res}
\ee
and corresponds to a binomial coefficient $C(L,N)$.
In this work, however, we are interested in a lattice model with multiply occupied sites.  
The partition function for this situation is 
\be
Z = \frac{(N+L-1)!}{N!(L-1)!}, 
\label{eq:Z_unres}
\ee
and corresponds to the binomial coefficient $C(N+L-1,N)$ and represents the permutation 
formula for $L-1$ items of type one and $N$ items of type two.  If $L-1$ items are assumed 
to represent bars, then these bars segregate $N$ items into $L$ sets, where a single set 
represents a lattice site.  Interpreted in this way, Eq. (\ref{eq:Z_unres}) makes perfect sense.  

Note, however, that 
the pressure per lattice site for the above system, defined as $P=\frac{\log Z}{\partial L}$, is 
\be
\beta P_{\rm} = \log\big(1+\rho\big), 
\label{eq:P_id}
\ee
and does not correspond to the ideal-gas behavior.  

To construct the partition function that reproduces ideal-gas properties, we must proceed from the
assumption that particles are distinguishable, in which case there are $L^N$ distinct configurations.  
The standard trick to obtain the partition function is to use the Gibbs correction, $1/N!$, yielding 
\be
Z = \frac{L^N}{N!}.  
\label{eq:Z_id}
\ee
The resulting partition function now yields the correct ideal-gas pressure per lattice site, 
\be
\beta P = \rho.  
\ee

Difference between the partition function in Eq. (\ref{eq:Z_unres}) and that in Eq. (\ref{eq:Z_id}) 
is well illustrated with different simulation algorithms.  One algorithm generates configurations 
by randomly selecting a lattice site.  This is followed by either adding or subtracting a particle.  
At the end of an update cycle, consisting of $L$ random picks of a lattice site, 
one ensures that the total number of particles in a system 
is conserved.  This algorithm corresponds to $Z$ in Eq. (\ref{eq:Z_unres}).  

In an alternative algorithm, configurations are generated by randomly selecting a particle (not a site),
hence, particles are labeled.  A selected particle is then moved to a randomly selected site.  
This algorithm corresponds to $Z$ in Eq. (\ref{eq:Z_id}) and is more suitable for representing liquids.

As it is more convenient to work with grand partition functions, below we obtain the 
corresponding expressions.  The formal relation between the grand and the canonical 
partition function is 
$$
\Xi(\beta \mu,L)=\sum_{N=0}^{\infty} e^{\beta\mu N} Z(N,L),
$$ 
where 
$\mu$ is the chemical potential and $\beta=1/k_BT$.  For indistinguishable particles, using 
Eq. (\ref{eq:Z_res}), $\Xi = \sum_{N=0}^{\infty} e^{\beta\mu N} \frac{(N+L-1)!}{N!(L-1)!}=(1-e^{\beta\mu})^{-L}$, 
where only $\mu<0$ is physically meaningful.  At $\mu=0$ the partition function diverges 
and for $\mu<0$ it becomes negative.  The same result is obtained from an alternative formulation 
\be
\Xi = \sum_{n_1=0}^{\infty}\dots \sum_{n_L=0}^{\infty}  e^{\beta\mu n_1}\dots e^{\beta\mu n_L}
= \bigg(\frac{1}{1-e^{\beta\mu}}\bigg)^L.  
\label{eq:Xi_unres}
\ee
The physical interpretation of the above expression is clear.  $L$ summations
correspond to $L$ sites and $n_i$ designates the number of particles at a site $i$. 
In the grand ensemble an average number of particles at each site is controlled with $\mu$.  

For distinguishable particles the grand partition function we use Eq. (\ref{eq:Z_unres}), leading to  
$\Xi = \sum_{N=0}^{\infty} e^{\beta\mu N} L^N/N!=(e^{e^{\beta\mu}})^L$.  The same result follows 
from an alternative formulation 
\be
\Xi = \sum_{n_1=0}^{\infty}\dots \sum_{n_L=0}^{\infty}  
\frac{e^{\beta\mu n_1}}{n_1!} \dots \frac{e^{\beta\mu n_L}}{n_L!}
= e^{Le^{\beta\mu}}.  
\label{eq:Xi_unres}
\ee

Later in this work we consider probabilities $p(n)$, the probability that any given site is 
occupied by $n$ particles.  Consequently, we derive these probabilities for non-interactive 
particles.  The procedures to obtain $p(n)$ will furthermore clarify the difference 
between distinguishable and indistinguishable particles.  

We start by recalling that $Z(N,L)$ 
in Eq. (\ref{eq:Z_unres}) counts the number of configurations for $N$ indistinguishable particles 
distributed over $L$ sites.  If one site is occupied by $n$ particles, the number of configurations 
of the remaining $N-n$ particles distributed over $L-1$ sites is $Z(N-n,L-1)$, and $p(n)$ is given
by the ratio of the two numbers, $p(n) = \frac{Z(N-n,L-1)}{Z(N,L)}$.  In the thermodynamic limit, 
$L\to\infty$ and $N\to\infty$ such that $N/L=\rho$ (and using the Sterling formula $N!\approx N^Ne^{-N}$ 
and the limiting representation of the exponential function $e^x = \lim_{N\to\infty}(1+x/N)^N$), that 
expression reduces to 
\be
p(n) = \frac{1}{1+\rho}\bigg(\frac{\rho}{1+\rho}\bigg)^n.  
\label{eq:pna}
\ee

For distinguishable particles the total number of configurations is $L^N$.  If one site is occupied
by $n$ labeled particles, the number of configurations of the remaining $N-n$ particles distributed 
over $L-1$ sites becomes $(L-1)^{N-n}$.  
However, the ratio $(L-1)^{N-n}/L^N$ does not yield the probability $p(n)$.  Since there are 
$\frac{N!}{n!(N-n)!}$ different ways to draw $n$ labeled particles, the ratio $(L-1)^{N-n}/L^N$ needs to 
be multiplied by that number.   
The correct distribution becomes 
\be
p(n) = \frac{N!}{n!(N-n)!} \frac{(L-1)^{N-n}}{L^N}, 
\label{eq:p_poisson}
\ee
which in the thermodynamic limit recovers the Poisson distribution
\be
p(n) = \frac{\rho^n e^{-\rho}}{n!}.  
\label{eq:pnb}
\ee

\section{Interactions:  one-component system}
\label{sec:model1}

We next consider an interactive 1d lattice system represented by the Hamiltonian 
\be
H(n_1,\dots,n_L) = \frac{K}{2} \sum_{i=1}^L n_i(n_i-1) +  \alpha K\sum_{i=1}^L n_in_{i+1}, 
\label{eq:H1D}
\ee
where the interactions between particles on the same site are given by the first term, 
and the interactions between particles on neighboring sites by the second term and are 
regulated by the dimensionless parameter $\alpha$.  For $\alpha=0$ 
interactions between particles on neighboring sites are turned off, and for $\alpha=1$ these
interactions are the same as those for particles on the same site.  The case $\alpha=1$
can be regarded as analogous to the Penetrable-Sphere model, and the case $0<\alpha<1$
to the Gaussian core model.  We are not interested in the scenario $\alpha>1$ which has no
correspondence in actual penetrable particle systems and implies that interactions between 
particles on neighboring sites are greater than those for particles on the same site.  
Finally, the scenario $\alpha<0$ implies attraction between particles on neighboring sites, 
in possible analogy to the van der Waals type of potentials, however, in this 
work we do not pursue this case.

As for non-interactive particles, we consider the system of indistinguishable and distinguishable
particles.  For indistinguishable particles the grand partition function is
\be
\Xi_a = \sum_{n_1=0}^{\infty}\dots\sum_{n_L=0}^{\infty} e^{\beta \mu n_1}\dots e^{\beta\mu n_L}  
e^{-\beta H(n_1,\dots,n_L)}, 
\label{eq:Za}
\ee
and for distinguishable particles it is
\be
\Xi_b = \sum_{n_1=0}^{\infty}\dots\sum_{n_L=0}^{\infty}  \frac{ e^{\beta \mu n_1} \dots 
e^{\beta \mu n_L} } {n_1!\dots n_L!}  e^{-\beta H(n_1,\dots,n_L)}, 
\label{eq:Zb}
\ee
where we use the index ``a'' and ``b'' to differentiate between the two cases.  Both cases adapt 
periodic boundary conditions, $n_{L+1}=n_1$, which ensures that each site is equivalent.

The systems are analyzed using the transfer matrix method, the standard method for 
dealing with lattice models in 1d \cite{Lavis}.  The central object of the method is the 
transfer matrix, $T(n,n')$, by means of which  
the partition function can be written as 
\be
\Xi = \sum_{n_1=0}^{\infty}\dots\sum_{n_L=0}^{\infty} T(n_1,n_2) T(n_2,n_3)\dots T(n_L,n_1), 
\label{eq:Z2}
\ee
revealing chain structure of a partition function.  Using matrix algebra, the partition function 
is shorthanded into 
\be
\Xi = {\rm Tr}\,{\bf T}^L.  
\ee
Eigendecomposition of the transfer matrix, ${\bf T}={\bf Q}{\bf \Lambda}{\bf Q}^{T}$ 
(where ${\bf \Lambda}$ is the diagonal matrix with diagonal elements $\Lambda_{ii}=\lambda_i$
and ${\bf Q}$ is the square matrix whose $i$-th column is the eigenvector $\phi_i(n)$ of ${\bf T}$), 
further transforms the expression into 
\be
\Xi = \sum_{i=1}^{\infty}\lambda_i^{L}, 
\label{eq:Xi_tm0}
\ee
where $\lambda_i^L$ are eigenvalues of the matrix ${\bf T}^L$, and ${\bf T}^L$ is the product matrix 
generated by multiplying ${\bf T}$ by itself $L$-times.  If eigenvalues are ordered 
according to their modulus as $|\lambda_1|>|\lambda_2|>|\lambda_3|\dots$, and because in 
the thermodynamic limit, $L\to\infty$, $\Xi$ is dominated by the largest eigenvalue, the grand partition 
function simply becomes  
\be
\Xi = \lambda_1^{L}, 
\label{eq:Xi_tm}
\ee
and the corresponding pressure per lattice site is given by  
\be
\beta P = \log \lambda_1. 
\label{eq:P1D} 
\ee

We next use the transfer matrix framework to obtain the probability $p(n)$, that a 
given site $i$ is occupied by $n$ particles.  The formal definition is 
\be
p(n) = \frac{1}{\Xi} \sum_{n_2=0}^{\infty}\dots\sum_{n_L=0}^{\infty} T(n,n_2) T(n_2,n_3)\dots T(n_L,n),  
\label{eq:p_n0}
\ee
and amounts to breaking the ring structure of Eq. (\ref{eq:Z2}) at a site $i=1$, 
giving rise to a linear chain.  
After the application of eigendecomposition (see Appendix \ref{sec:A1}), 
the expression reduces to 
\be
p(n) = \phi_1^2(n),  
\label{eq:p_n}
\ee
where $\phi_1(n)$ are elements of the dominant eigenvector corresponding to the 
largest eigenvalue $\lambda_1$, thus, $p(n)$ is properly normalized since the modulus of 
a vector $\phi_1(n)$ is $1$.

The transfer matrix for indistinguishable particles, corresponding to the partition function in 
Eq. (\ref{eq:Za}), is
\be
T_a(n,n') = e^{-\frac{\beta K}{4}(n^2+n'^2)} e^{-\beta \alpha K n n'} \,e^{\frac{\beta \mu'}{2} (n+n') } , 
\label{eq:Ta}
\ee
and that for distinguishable particles corresponding to $Z$ in Eq. (\ref{eq:Zb}) is
\be
T_b(n,n') = e^{-\frac{\beta K}{4}(n^2+n'^2)}  e^{-\beta \alpha K n n'} \,\frac{e^{\frac{\beta \mu'}{2} (n+n') }}{\sqrt{n!n'!}}.  
\label{eq:Tb}
\ee
Eignevalues and the eigenvectors are calculated numerically using Mathematica.  In principle, 
$T(n,n')$ is an infinite matrix, but in practice the $20\times 20$ matrix is sufficient for most 
situations.  

\begin{figure}[h] 
 \begin{center}
 \begin{tabular}{rrrr}
  \includegraphics[height=0.15\textwidth,width=0.16\textwidth]{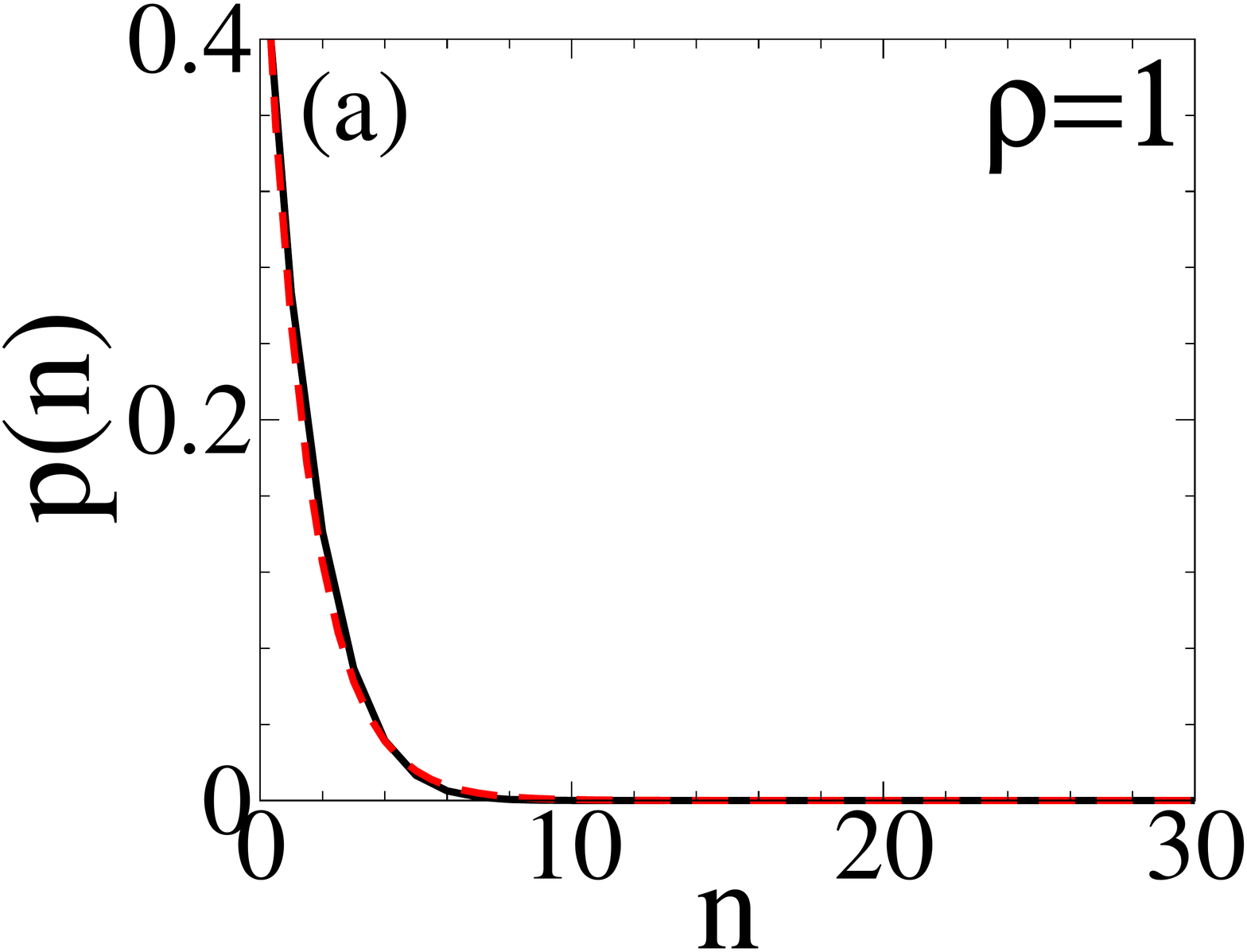}&
\hspace{-0.25cm}  
  \includegraphics[height=0.15\textwidth,width=0.16\textwidth]{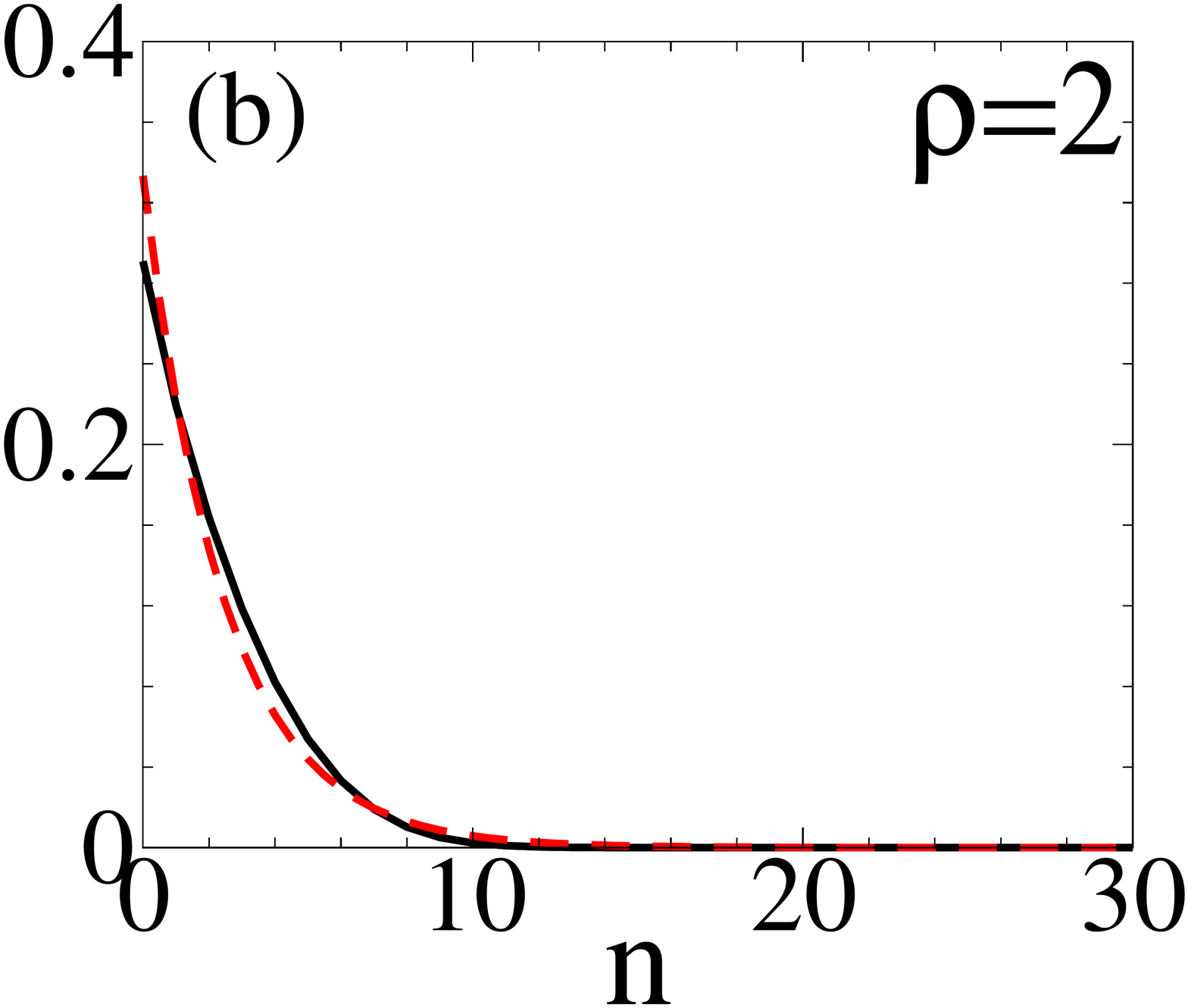}&
\hspace{-0.25cm}  
  \includegraphics[height=0.15\textwidth,width=0.16\textwidth]{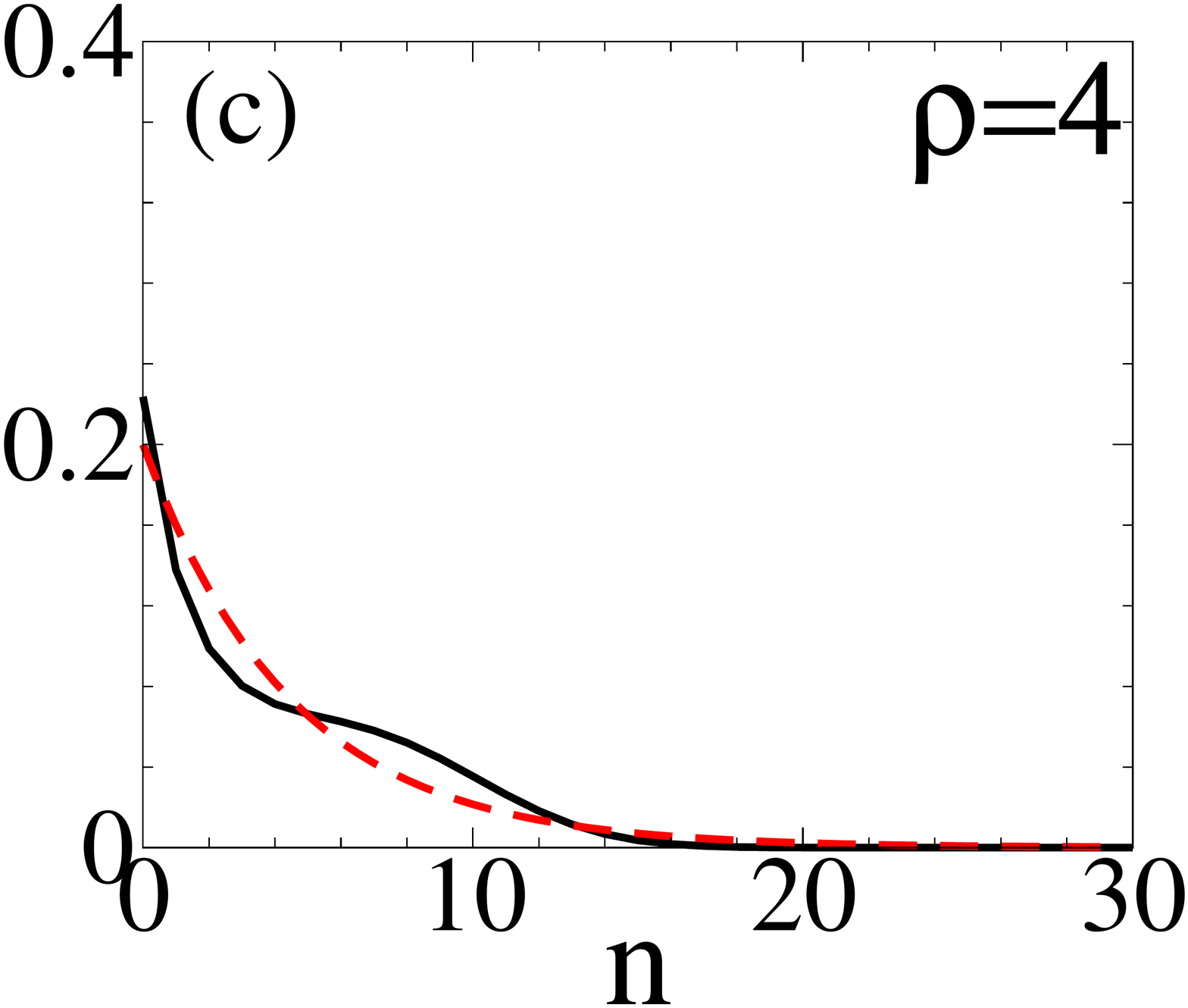}\\
  \includegraphics[height=0.15\textwidth,width=0.16\textwidth]{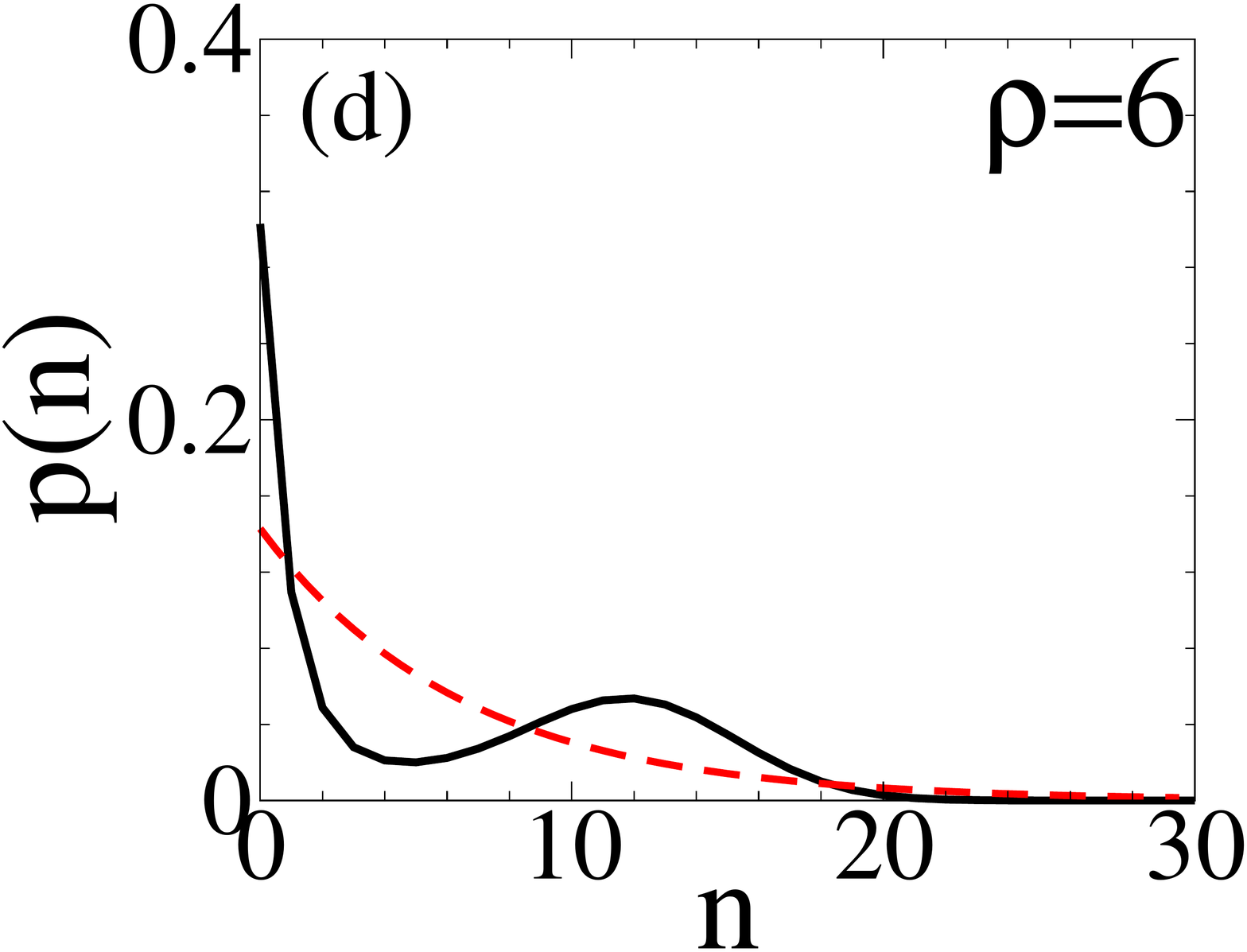}&
\hspace{-0.25cm}
  \includegraphics[height=0.15\textwidth,width=0.16\textwidth]{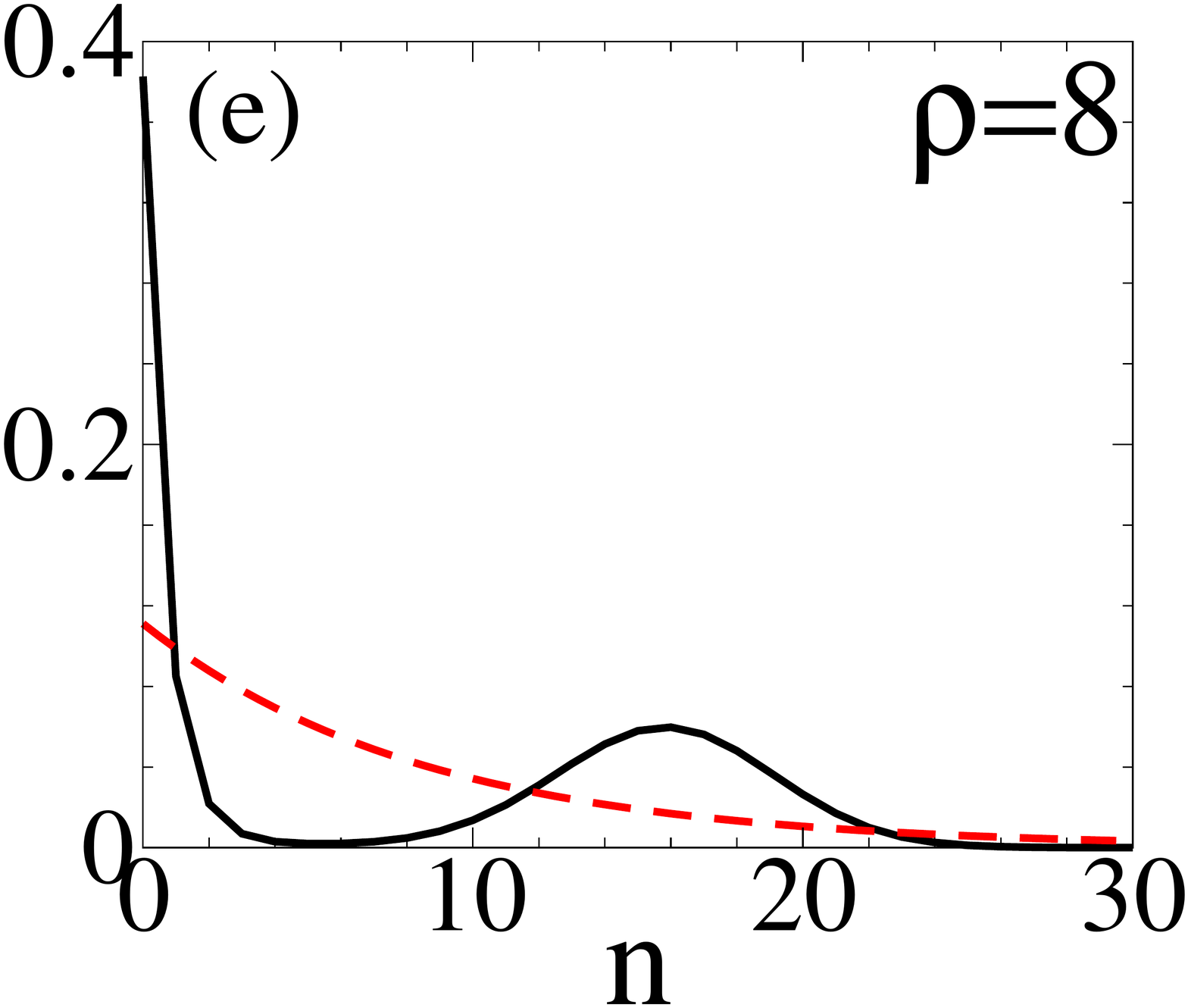}&
\hspace{-0.25cm}
  \includegraphics[height=0.15\textwidth,width=0.16\textwidth]{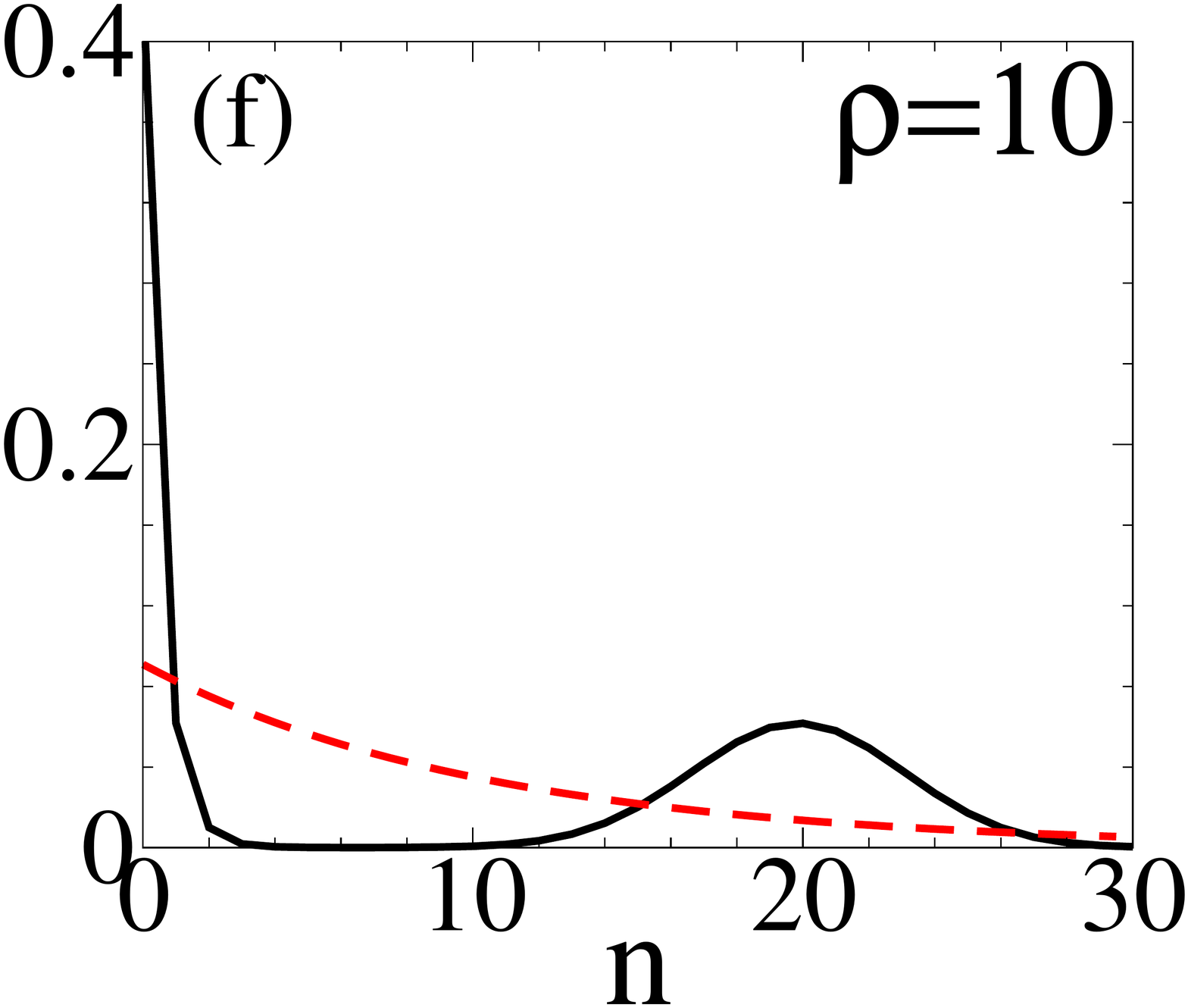}&
 \end{tabular}
 \end{center}
\caption{$p(n)$ for indistinguishable particles for $\beta K=0.1$ and $\alpha=1$.  
The dashed red line is for non-interactive particles according to the expression in 
Eq. (\ref{eq:pna}).}
\label{fig:p1} 
\end{figure}

Fig. (\ref{fig:p1}) shows a number of distributions $p(n)$ for indistinguishable particles, for $\beta K=0.1$ 
and $\alpha=1$, for different densities arranged in increasing order.  The consecutive 
plots show gradual transformation of $p(n)$ into a bimodal structure, emerging at around $\rho = 4$.  
The two peaks of the bimodal structure are at $n=0$ and $n\approx 2\rho$, suggesting an alternating 
structure of occupied versus empty sites, rather than a coexistence of 
vacuum cavities embedded in a fluid with density $2\rho$.  
The emergence of an alternating structure is analogous to the ``stack'' formations  of penetrable-spheres
discussed in the introduction.  
\graphicspath{{figures/}}

\begin{figure}[h] 
 \begin{center}
 \begin{tabular}{rrrr}
  \includegraphics[height=0.15\textwidth,width=0.16\textwidth]{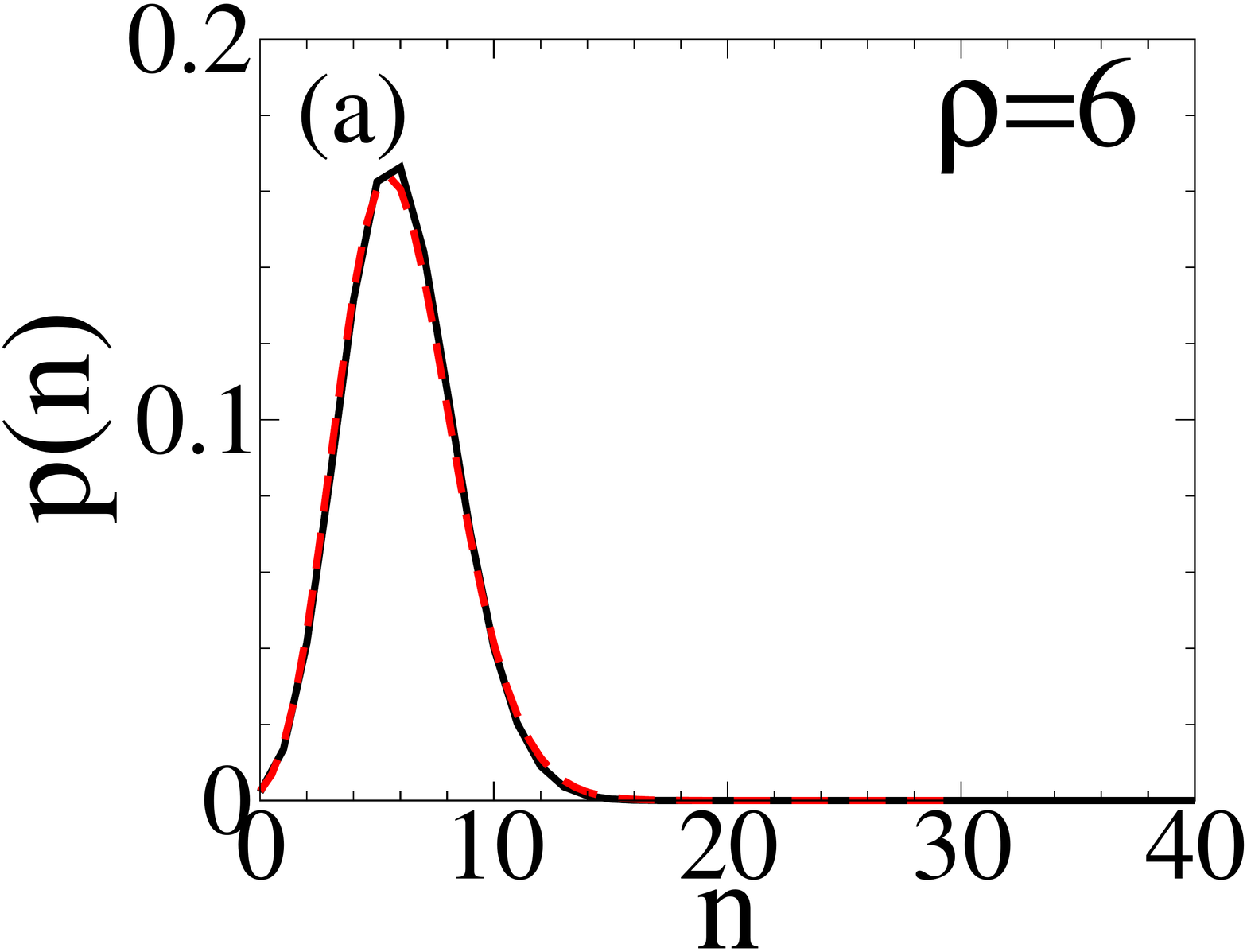}&
\hspace{-0.25cm}
  \includegraphics[height=0.15\textwidth,width=0.16\textwidth]{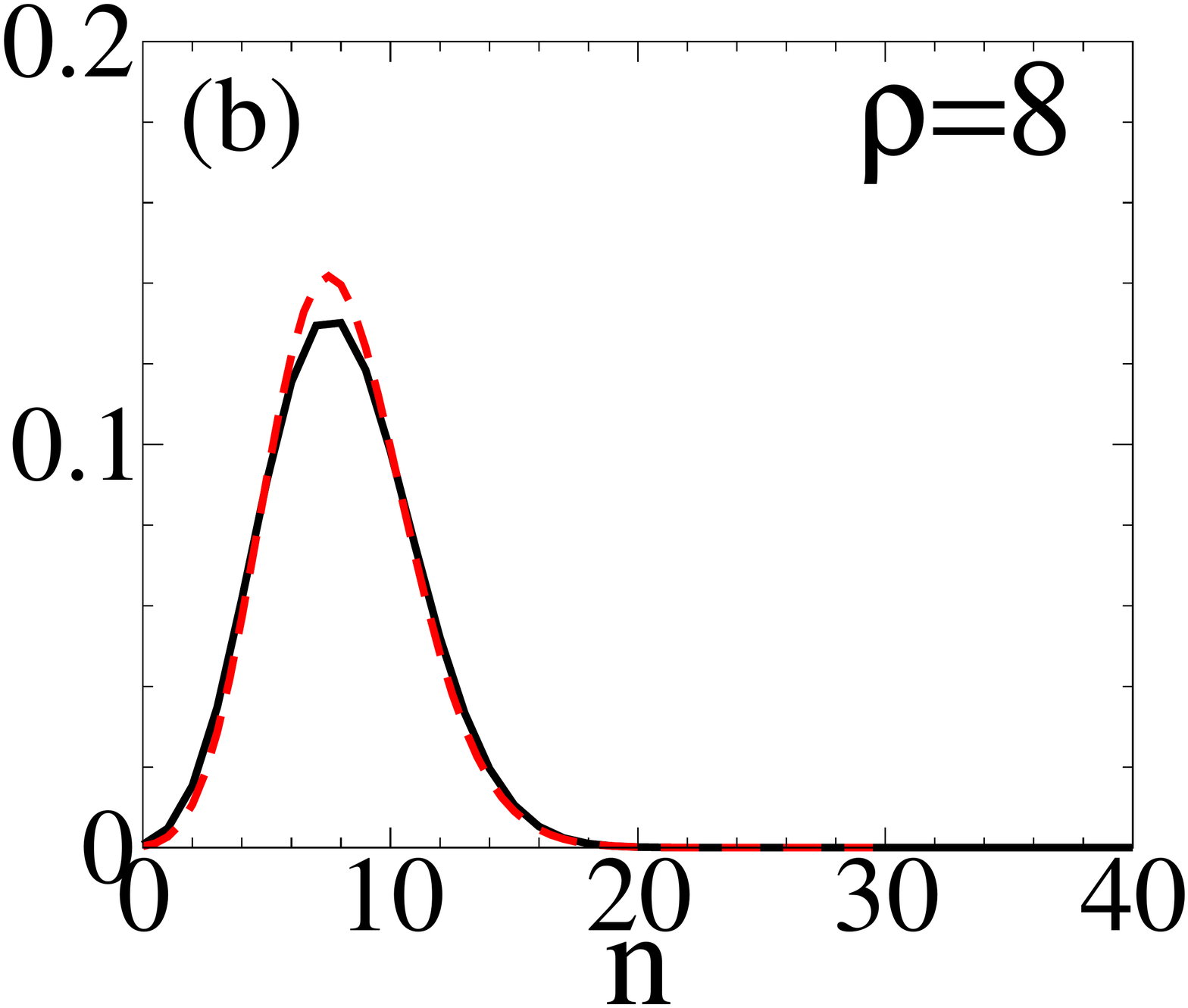}&
\hspace{-0.25cm}
  \includegraphics[height=0.15\textwidth,width=0.16\textwidth]{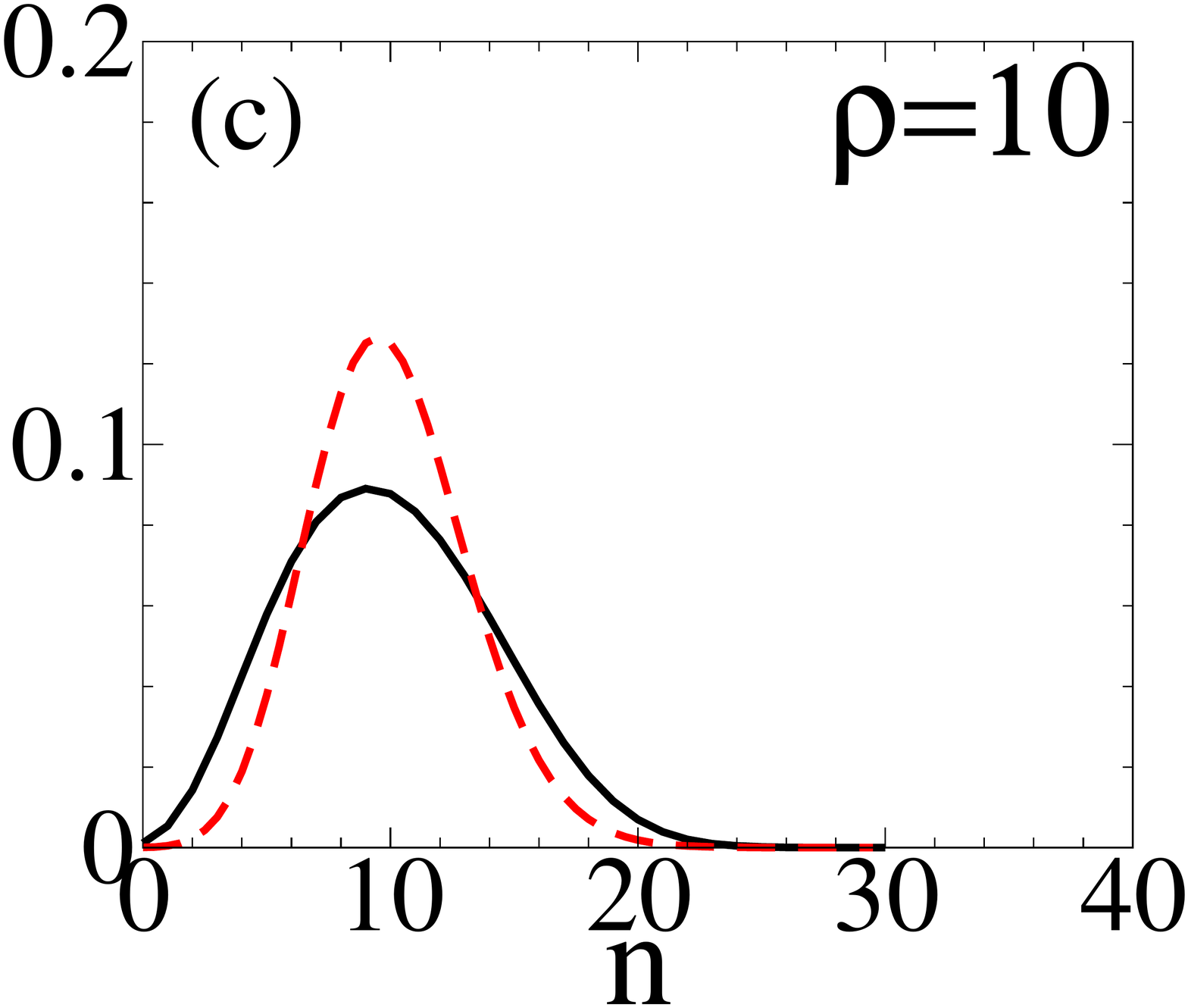}\\
  \includegraphics[height=0.15\textwidth,width=0.16\textwidth]{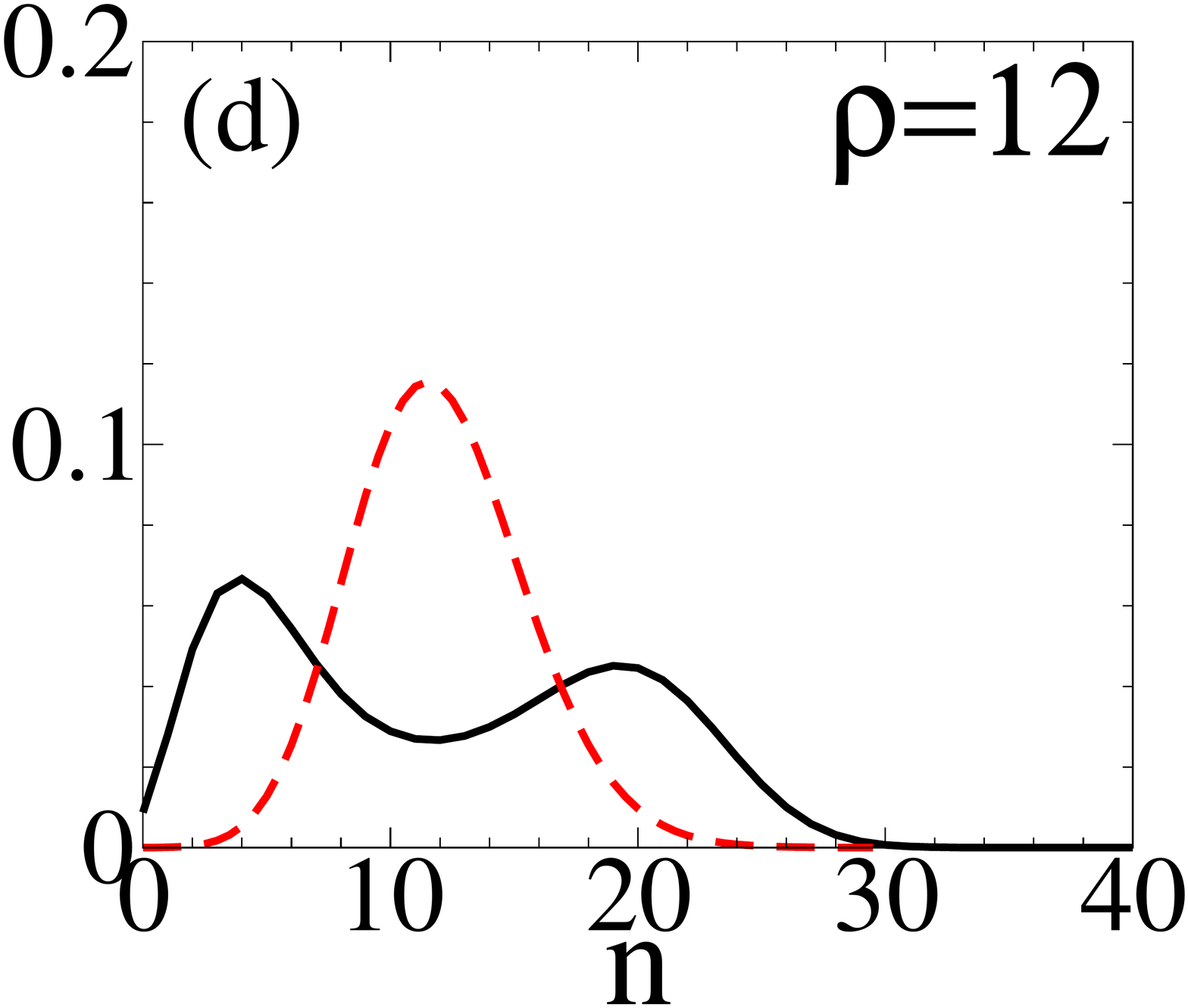}&
\hspace{-0.25cm}
  \includegraphics[height=0.15\textwidth,width=0.16\textwidth]{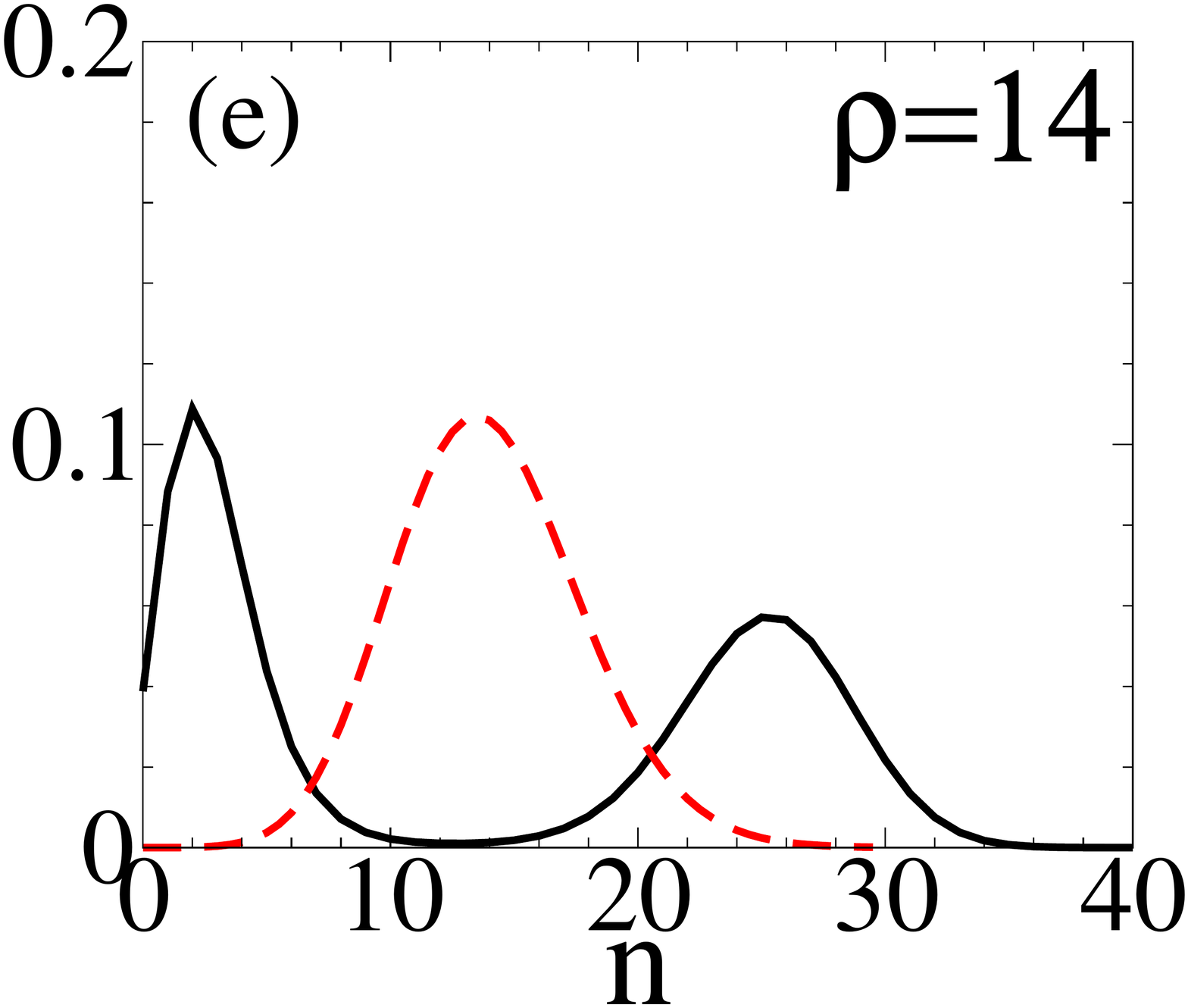}&
\hspace{-0.25cm}
  \includegraphics[height=0.15\textwidth,width=0.16\textwidth]{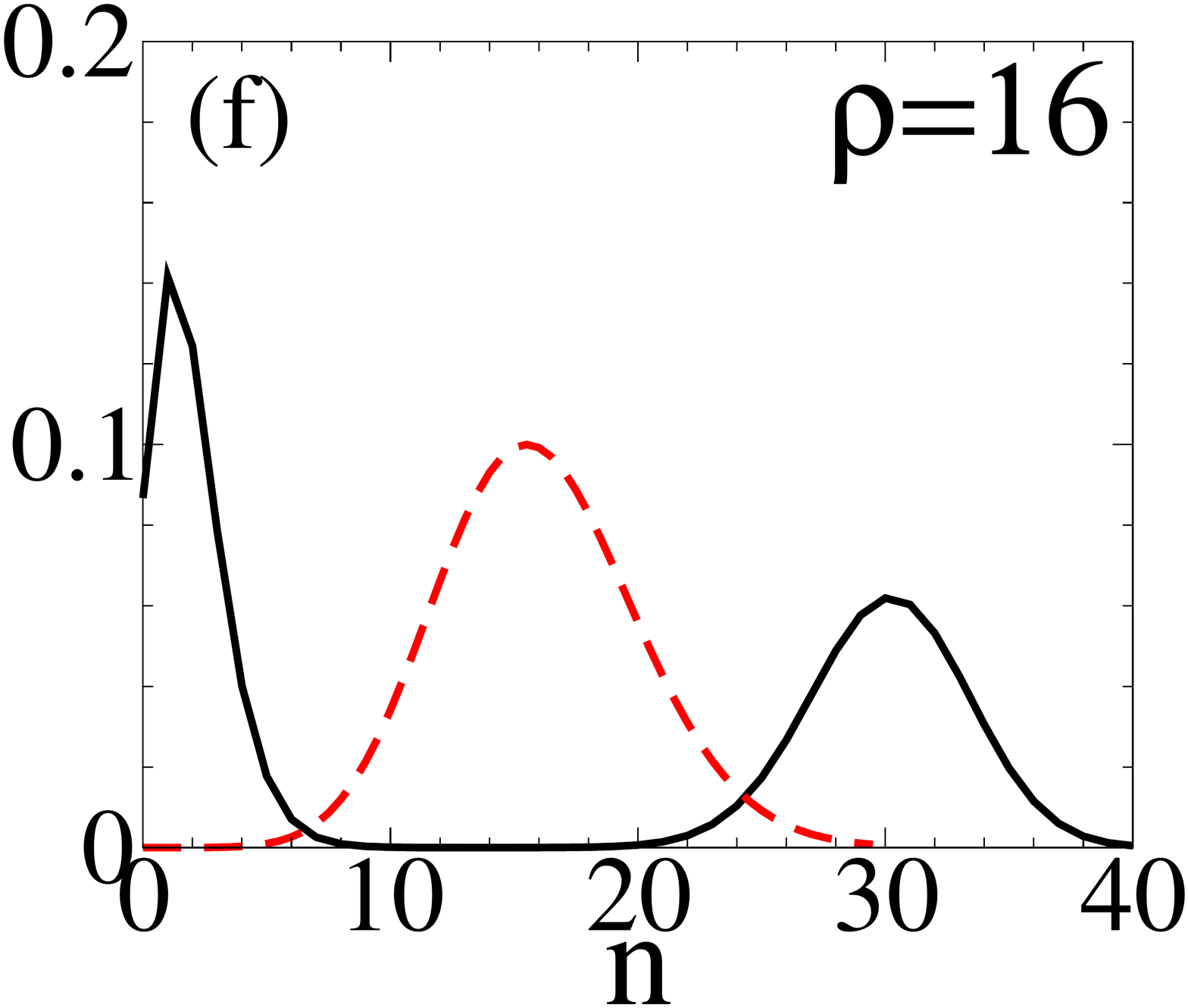}&
 \end{tabular}
 \end{center}
\caption{$p(n)$ for distinguishable particles  for $\beta K=0.1$ and $\alpha=1$.  
The dashed red line corresponds to the Poisson distribution of non-interactive 
particles in Eq. (\ref{eq:pnb}).}
\label{fig:p2} 
\end{figure}

A similar transformation into a bimodal structure occurs for distinguishable particles, 
see Fig. (\ref{fig:p2}).  The crossover, however, occurs at a higher density, $\rho\approx 10$.  
The explanation for this difference lies in different entropies of the two systems.  
For the case of distinguishable particles the adaptation of an ordered 
alternating structure entails larger loss of entropy.  
\graphicspath{{figures/}}

To confirm the existence of an alternating structure, we consider the two-site probability, 
$p_m(n,n')$, which is the probability that two sites separated by $m$ sites have the occupation 
number $n$ and $n'$, and whose formal definition is 
\ba
p_m(n,n') &=& \frac{1}{\Xi} \sum_{n_2=0}^{\infty}\dots\sum_{n_{m}=0}^{\infty}T(n,n_2) T(n_2,n_3)\dots T(n_m,n') \nonumber\\
&&\dots \sum_{n_{m+2}=0}^{\infty}\dots \sum_{n_L=0}^{\infty} T(n',n_{m+2})\dots T(n_L,n). \nonumber\\ 
\ea
Applying eigendecomposition (see Appendix \ref{sec:A2}), we arrive at the expression
\ba
\frac{p_m(n,n')}{p(n)p(n')} =  1 
+ \sum_{k=2}^{\infty} \bigg(\frac{\lambda_k}{\lambda_1}\bigg)^m 
\frac{\phi_k(n)  \phi_k(n')}{\phi_1(n)  \phi_1(n')}, \nonumber\\
\label{eq:p12}
\ea
where $p_m(n,n')$ can be shown to be normalized, 
\ba
\sum_{n=0}^{\infty}\sum_{n'=0}^{\infty} p_m(n,n') &=& \sum_{n=0}^{\infty}p(n)\sum_{n'=0}^{\infty} p(n') \nonumber\\
&&\!\!\!\!\!\!\!\!\!\! \!\!\!\!\!\!\!\!\!\! \!\!\!\!\!\!\!\!\!\! \!\!\!\!\!\!\!\!\!\! \!\!\!\!\!\!\!\!\!\! 
+ \sum_{k=2}^{\infty} \bigg(\frac{\lambda_k}{\lambda_1}\bigg)^m\sum_{n=0}^{\infty}  \phi_1(n) \phi_k(n)  
\sum_{n'=0}^{\infty}\phi_1(n')\phi_k(n') = 1, \nonumber\\ 
\ea
where the second term vanishes for any $k\ne 1$ 
as the consequence of orthonormality of the eigenvectors $\phi_k$.  

We next define the quantity 
\be
\Gamma_m = \frac{p_m(0,0)}{p^2(0)} - 1.  
\label{eq:Gamma}
\ee
In absence of correlations between empty sites, $\Gamma_m=0$.  On the other 
hand, if occupied and empty sites alternate, we expect 
\be
  \begin{array}{r l}
       \Gamma_m > 0, & \quad \text{for $m$-even}\\
       \Gamma_m < 0, & \quad \text{for $m$-odd}.  
  \end{array} 
  \ee

\begin{figure}[h] 
 \begin{center}
 \begin{tabular}{rrrr}
  \includegraphics[height=0.18\textwidth,width=0.22\textwidth]{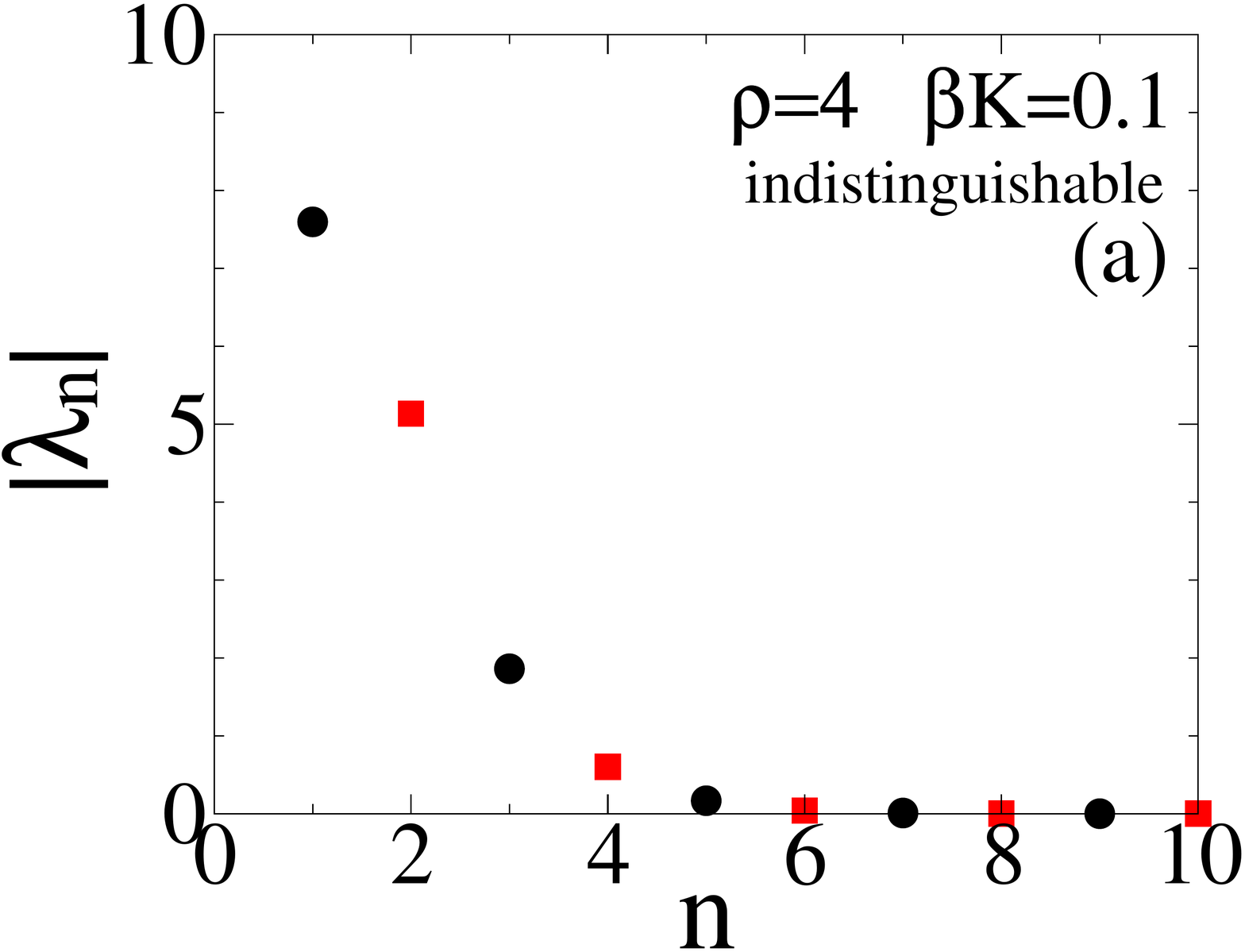}&
  \includegraphics[height=0.18\textwidth,width=0.22\textwidth]{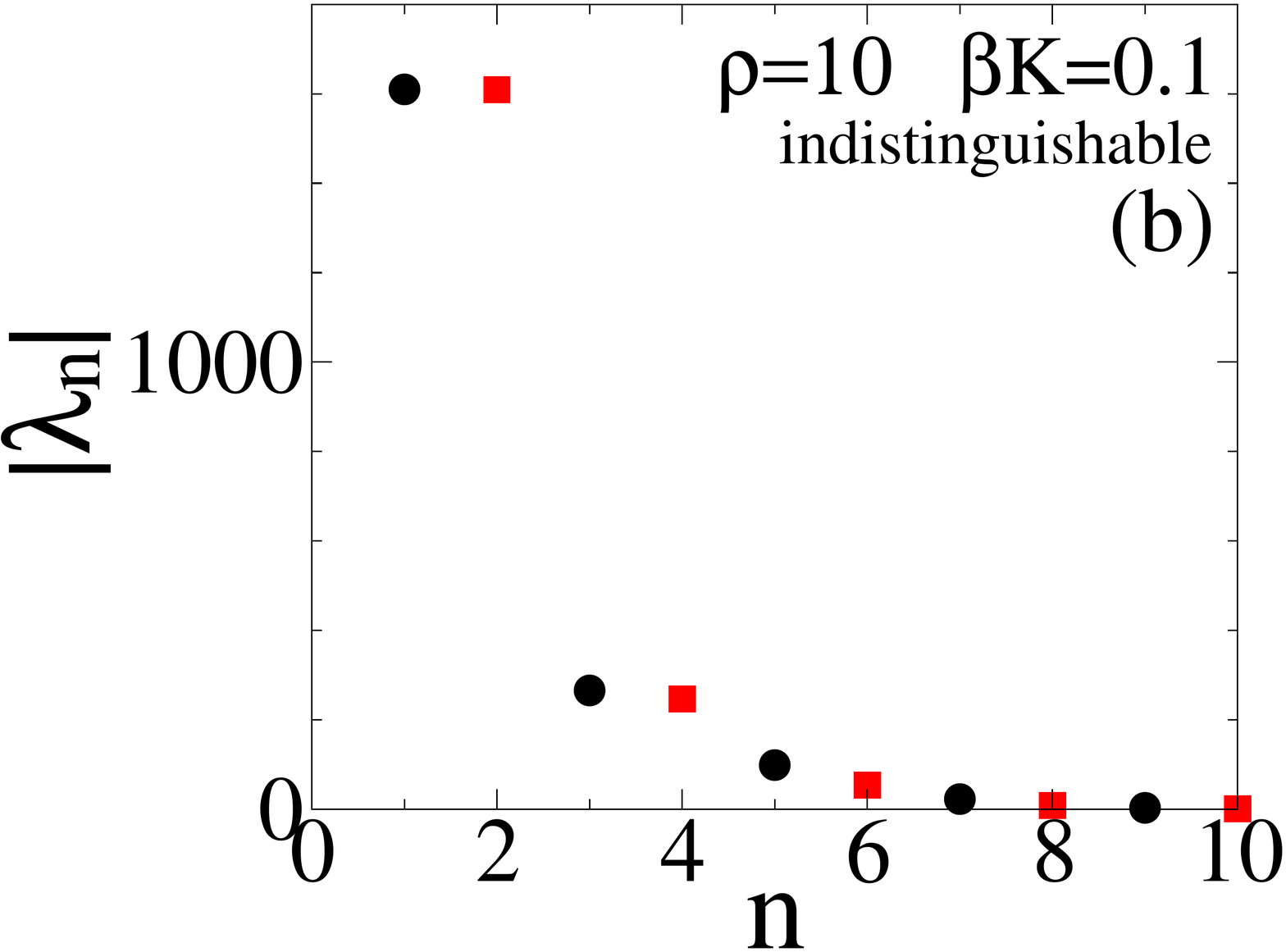}&
 \end{tabular}
 \end{center}
\caption{Partially ordered eigenvalues $|\lambda_1|>|\lambda_2|>\dots$, for  for $\beta K=0.1$ and $\alpha=1$.  
Circles are for $n$ odd, corresponding to $\lambda_n$ positive, and squares are for $n$ even, 
corresponding to $\lambda_n$ negative. }
\label{fig:lambda} 
\end{figure}

\begin{figure}[h] 
 \begin{center}
 \begin{tabular}{rrrr}
  \includegraphics[height=0.19\textwidth,width=0.23\textwidth]{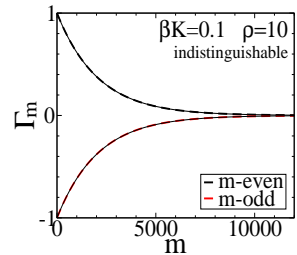}&
 \end{tabular}
 \end{center}
\caption{The correlation function $\Gamma_m$ defined in Eq. (\ref{eq:Gamma}) as 
a function of $m$, for $\beta K=0.1$ and $\alpha=1$.  The thick dashed lines are exact
results, and the thin solid lines are for the ansatz $\pm e^{-m\log(\lambda_1/\lambda_2)}$.}
\label{fig:gamma_m} 
\end{figure}

In Fig. (\ref{fig:lambda}) we plot $|\lambda_n|$ for the density when the distribution
starts to separate into bimodal structure and for the density where the distribution has a well developed 
bimodal structure.  The eigenvalues alternate in sign as 
\be
  \begin{array}{r l}
       \lambda_n > 0, & \quad \text{for $n$-odd}\\
       \lambda_n < 0, & \quad \text{for $n$-even},
  \end{array} 
  \ee
which is not captured by the figure which plots the data points for $|\lambda_n|$.  
The main result is that once the bimodal structure is established, the spectrum is 
dominated by the first two eigenvalues, $\lambda_1$ and $\lambda_2$, so that $\Gamma_m$
in Eq. (\ref{eq:Gamma}) can be approximated as 
\be
\Gamma_m \approx \bigg(\frac{\lambda_2}{\lambda_1}\bigg)^m \bigg(\frac{\phi_2(0)}{\phi_1(0)}\bigg)^2.  
\ee
Furthermore, we find that $\phi_2(0)/\phi_1(0) \approx 1$, so that 
the correlations are determined solely by the ratio $\lambda_2/\lambda_1<0$.  Since $\lambda_1$
and $\lambda_2$ have different sign, $\lambda_2/\lambda_1$ raised 
to odd power is negative, and raised to even power it is positive.  This suggests  
\be
  \begin{array}{r l}
       \Gamma_m \approx \,\,\,\,\, e^{-m\log(\lambda_1/\lambda_2)}, & \quad \text{for $m$-even}\\
       \Gamma_m \approx -e^{-m\log(\lambda_1/\lambda_2)}, & \quad \text{for $m$-odd}. 
  \end{array} 
  \label{eq:Gamma_2}
  \ee
Fig. (\ref{fig:lambda}) indicates that the first two eigenvalues converge,  
$|\lambda_2|\to |\lambda_1|$, but attain equality only in the limit $\rho\to\infty$.  
Since the condition $|\lambda_2|= |\lambda_1|$ implies long-range order, 
the phase transition in 1d does not take place at finite density \cite{Cuesta04}.  
Only for dimensionality $d>1$ a phase transition is possible.  
\graphicspath{{figures/}}
$\Gamma_m$ plotted in Fig. (\ref{fig:gamma_m}) confirms the exponentially decaying correlations
and accuracy of the ansatz in Eq. (\ref{eq:Gamma_2}).  
\graphicspath{{figures/}}

So far we have considered only the case $\alpha=1$, which can be regarded as representative 
of the Penetrable-Sphere model whose interaction strength remains the same as long as spheres 
are overlaped.  In the subsequent section we look into other values of $\alpha$, especially, 
we examine the effect of reduced $\alpha$ on the structure of $p(n)$.  

\begin{figure}[h] 
 \begin{center}
 \begin{tabular}{rrrr}
  \includegraphics[height=0.15\textwidth,width=0.16\textwidth]{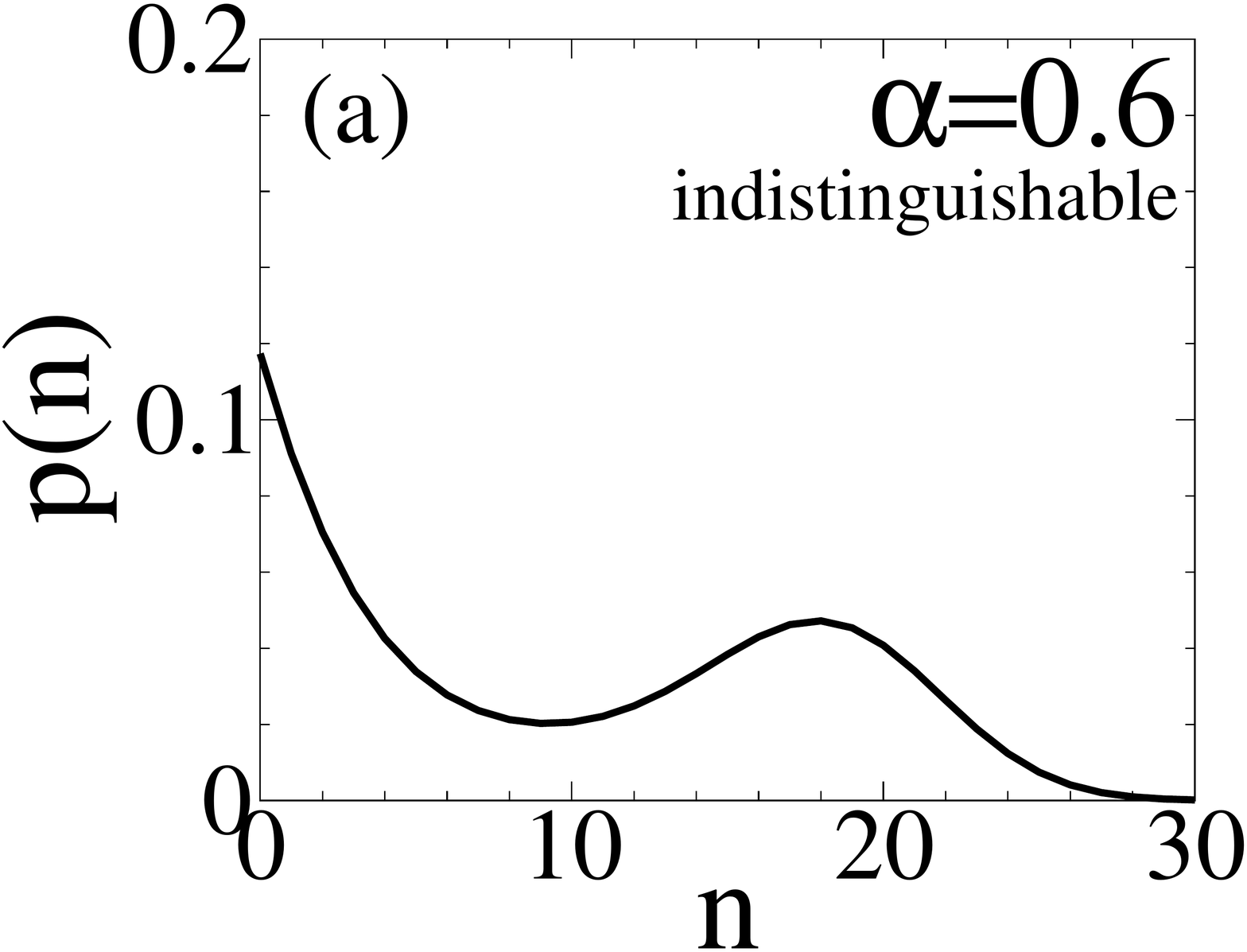}&
\hspace{-0.25cm}
  \includegraphics[height=0.15\textwidth,width=0.16\textwidth]{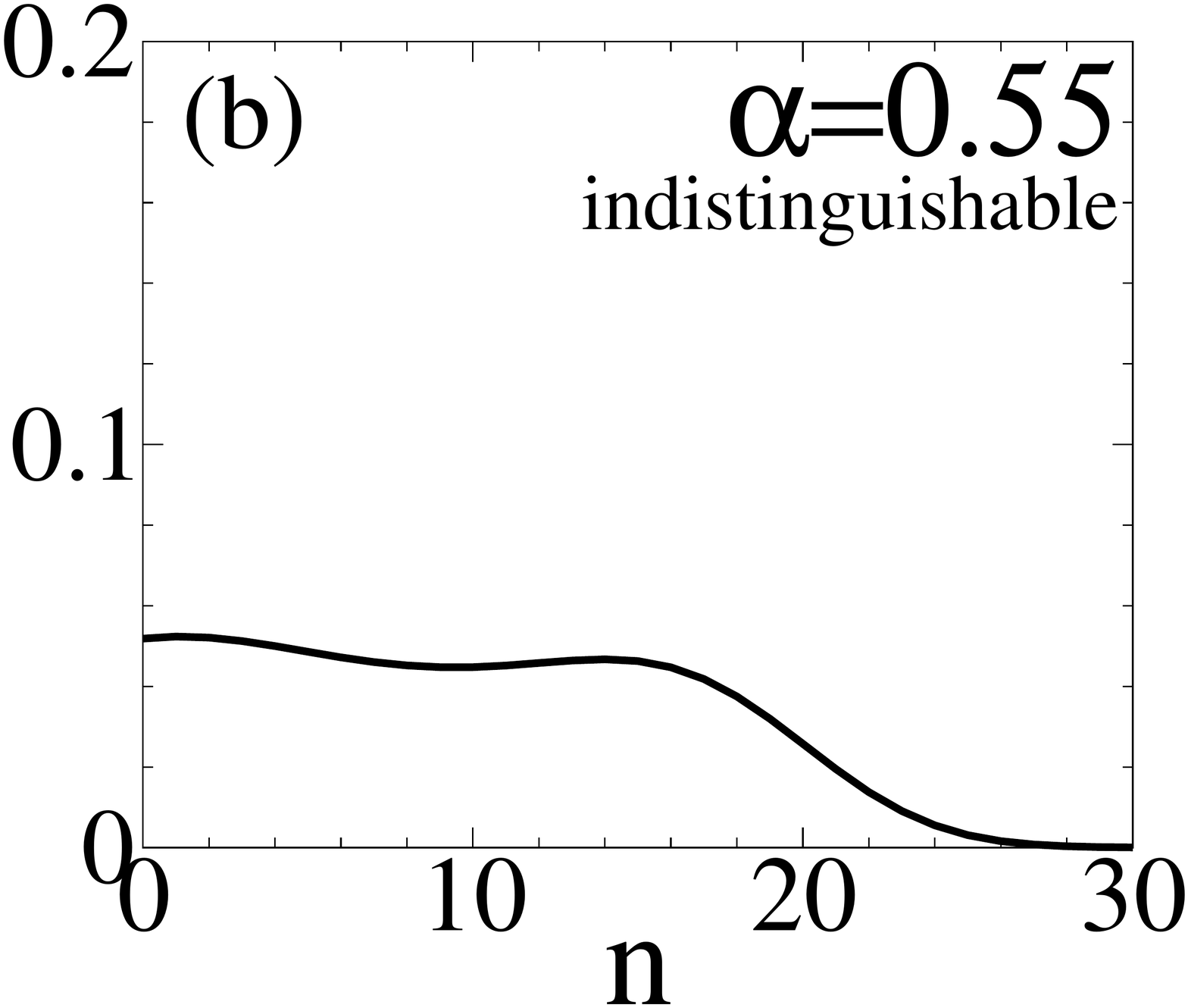}&
\hspace{-0.25cm}
  \includegraphics[height=0.15\textwidth,width=0.16\textwidth]{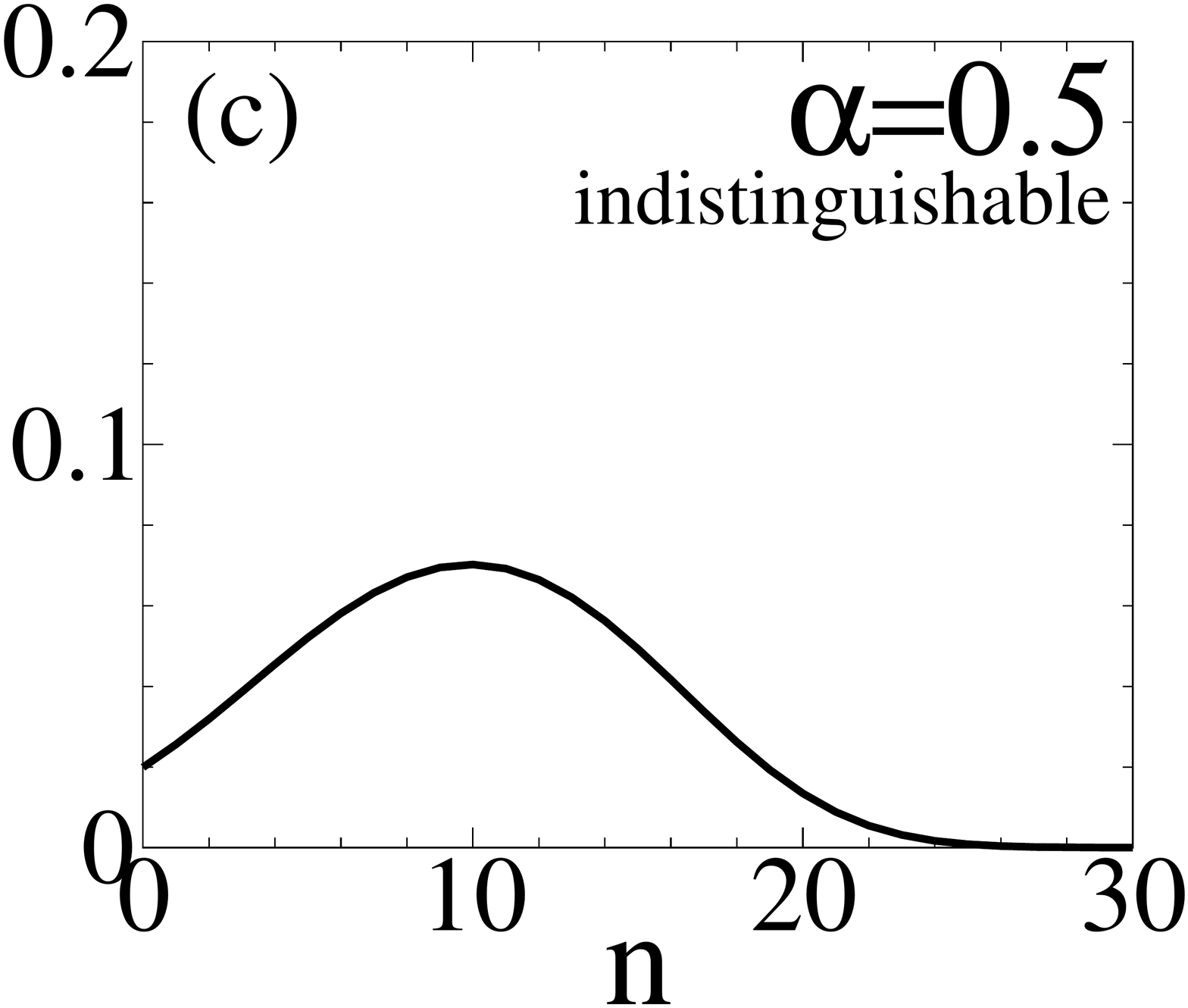}\\
  \includegraphics[height=0.15\textwidth,width=0.16\textwidth]{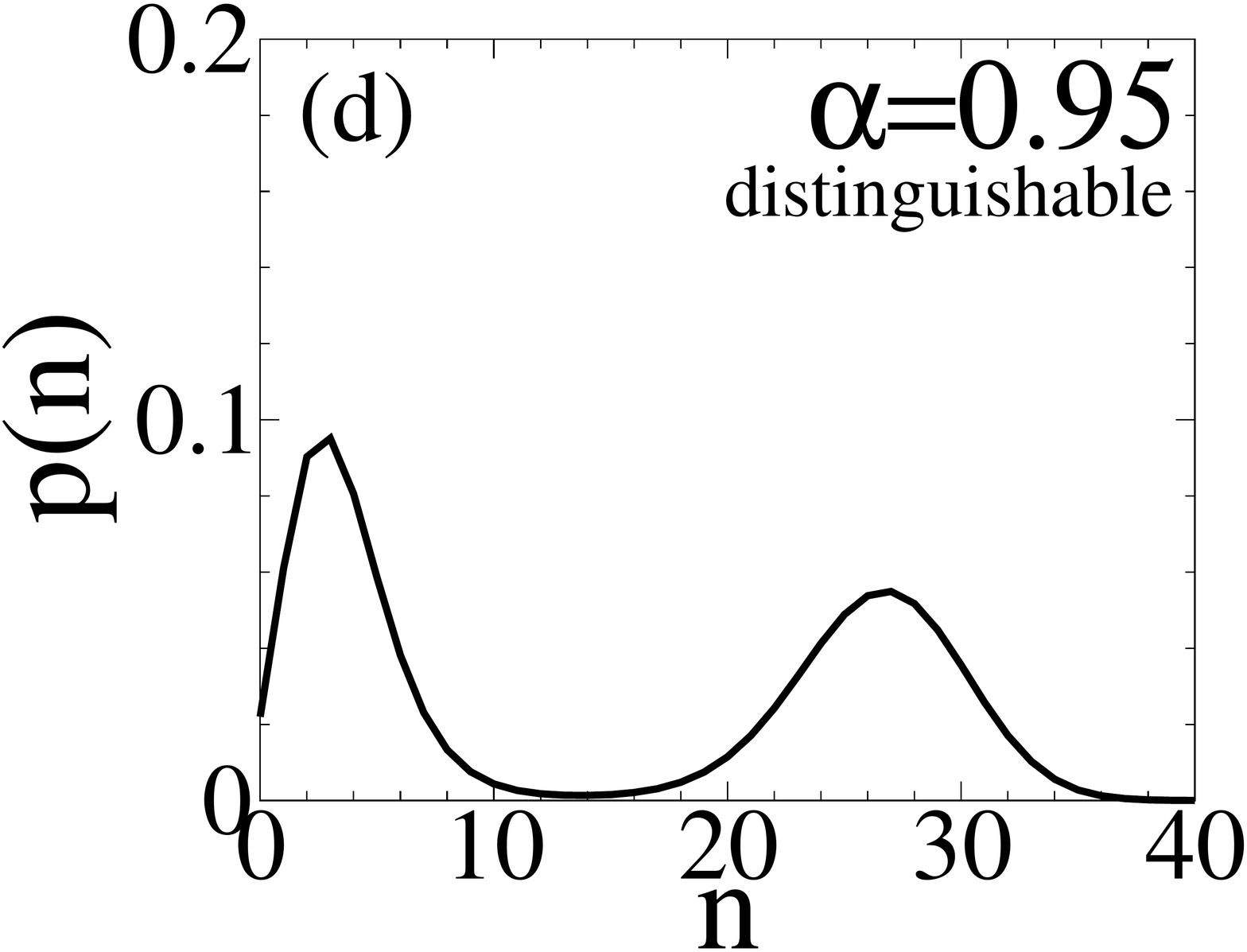}&
\hspace{-0.25cm}
  \includegraphics[height=0.15\textwidth,width=0.16\textwidth]{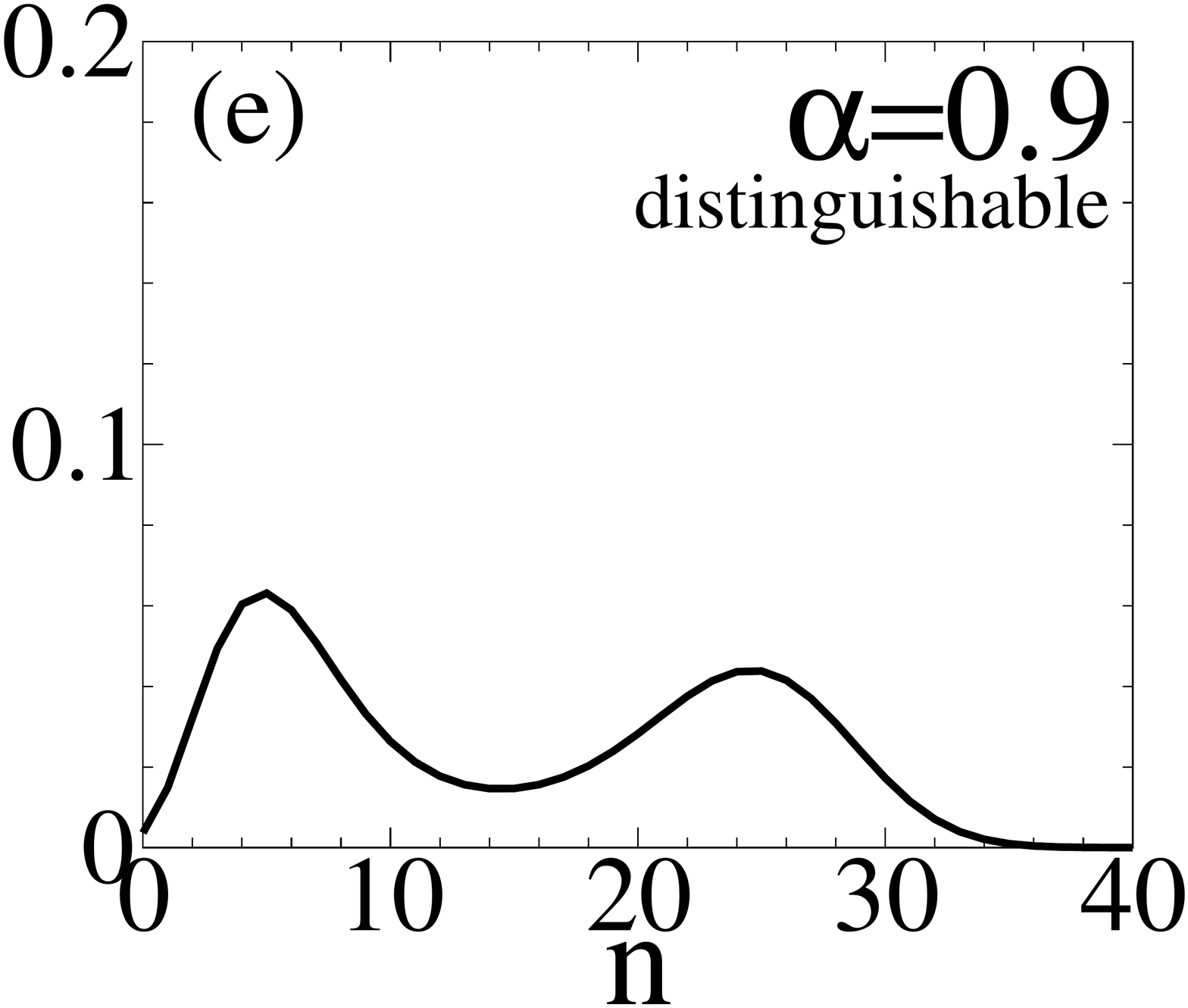}&
\hspace{-0.25cm}
  \includegraphics[height=0.15\textwidth,width=0.16\textwidth]{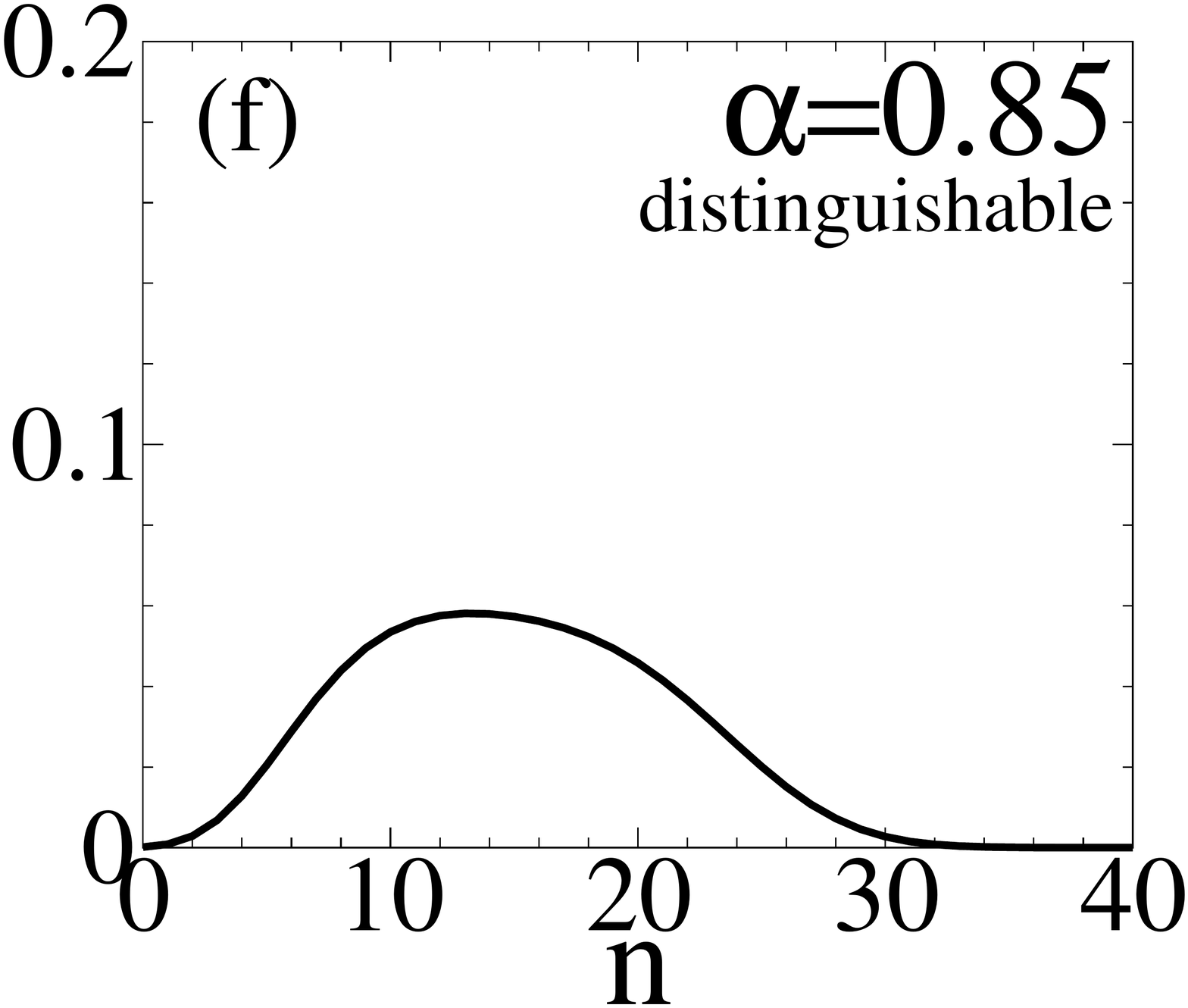}&
 \end{tabular}
 \end{center}
\caption{$p(n)$ for different $\alpha$.  The interaction strength is $\beta K=0.1$.  
The density for indistinguishable particles $\rho=10$ and that for distinguishable particles 
it is $\rho=15$.}
\label{fig:p3} 
\end{figure}

In Fig. (\ref{fig:p3}) we plot several distributions 
$p(n)$, for $\beta K=0.1$ and fixed $\rho$, for decreasing values of $\alpha$.  The results 
indicate gradual transformation of a bimodal into a mono-modal structure, implying 
the dissolution of an ordered alternating structre. For indistinguishable particles at $\alpha=0.55$ the 
bimodal structure is no longer there, however, it reappears if density is increased.  
The question is, what is the critical 
value of $\alpha$ below which the ordered alternating structure never arises for any density?
\graphicspath{{figures/}}

To answer this, we consider two idealized configurations.  One configuration is uniform, with 
each site having the occupation number $n_i=\rho$.  Another one have alternating occupations 
between $n_i=2\rho$ and $n_i=0$.  From the Hamiltonian in Eq. (\ref{eq:H1D}), energy of each 
configuration is 
\be
E_{\rm hom} = \frac{LK}{2} \bigg[(1+2\alpha)\rho^2 - \rho\bigg], 
\label{eq:E1}
\ee 
and 
\be
E_{\rm alt} =  \frac{LK}{2} \bigg[2\rho^2 - \rho\bigg], 
\label{eq:E2}
\ee 
and the energy gained by each site by replacing a homogenous structure with an alternating structure is 
\be
\frac{\Delta E}{L}  =  -\frac{K\rho^2}{2} (2\alpha-1).  
\label{eq:DE}
\ee
The results indicate that for $\alpha\le 0.5$ there is no longer an energy gain by adapting 
an ordered alternating structure.   
The condition $\alpha>0.5$, however, is sufficient but not necessary to have an alternating 
structure, as this condition was determined from energy considerations alone.  In real systems the 
energy gain, by switching to more ordered structure, is accompanied 
by loss of entropy.  In addition to the condition $\alpha>0.5$, 
we need the condition where energy is a dominant contribution of the 
free energy.  This is attained for large interaction strength and large density.  

\begin{figure}[h] 
 \begin{center}
 \begin{tabular}{rrrr}
  \includegraphics[height=0.18\textwidth,width=0.22\textwidth]{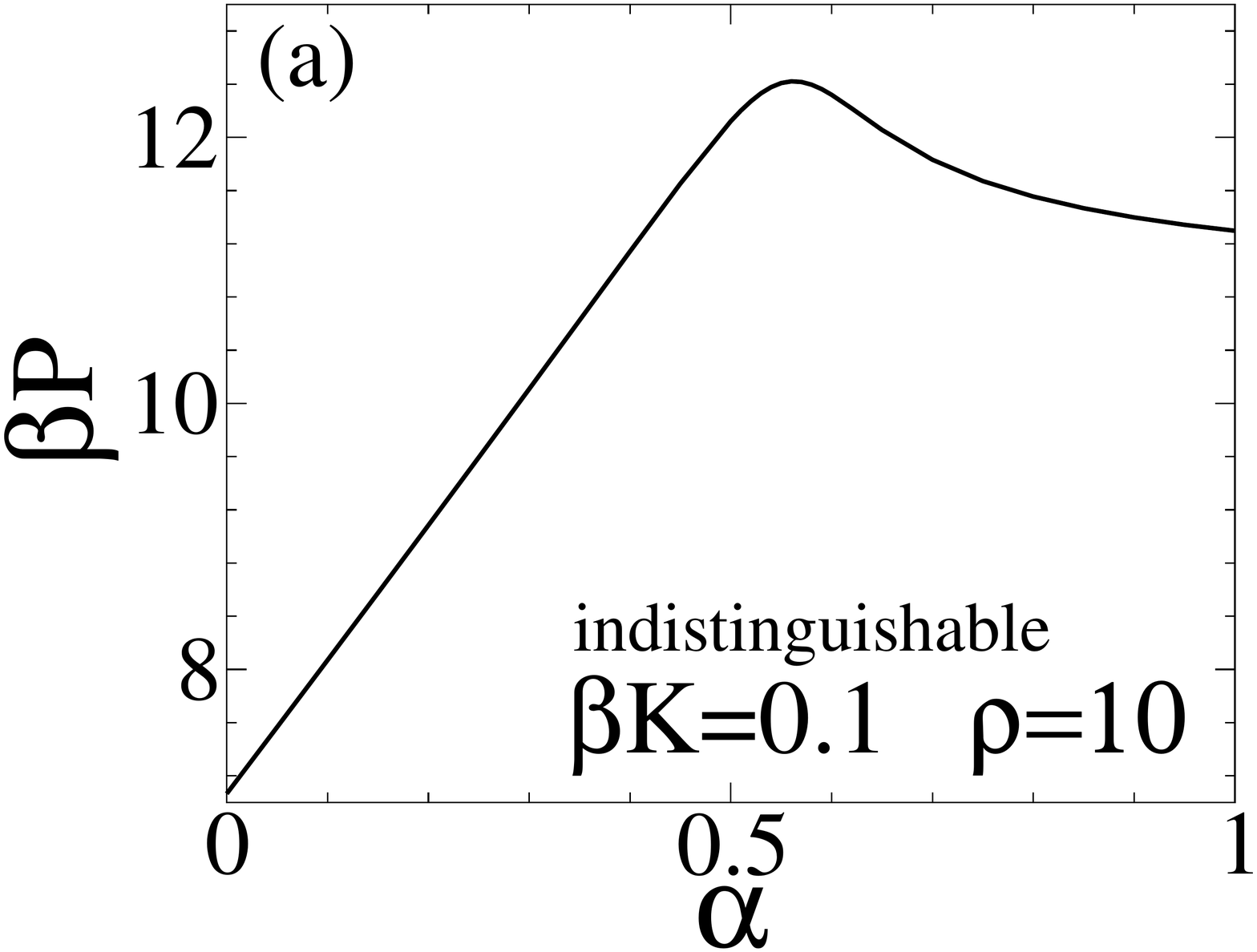}&
  \includegraphics[height=0.18\textwidth,width=0.22\textwidth]{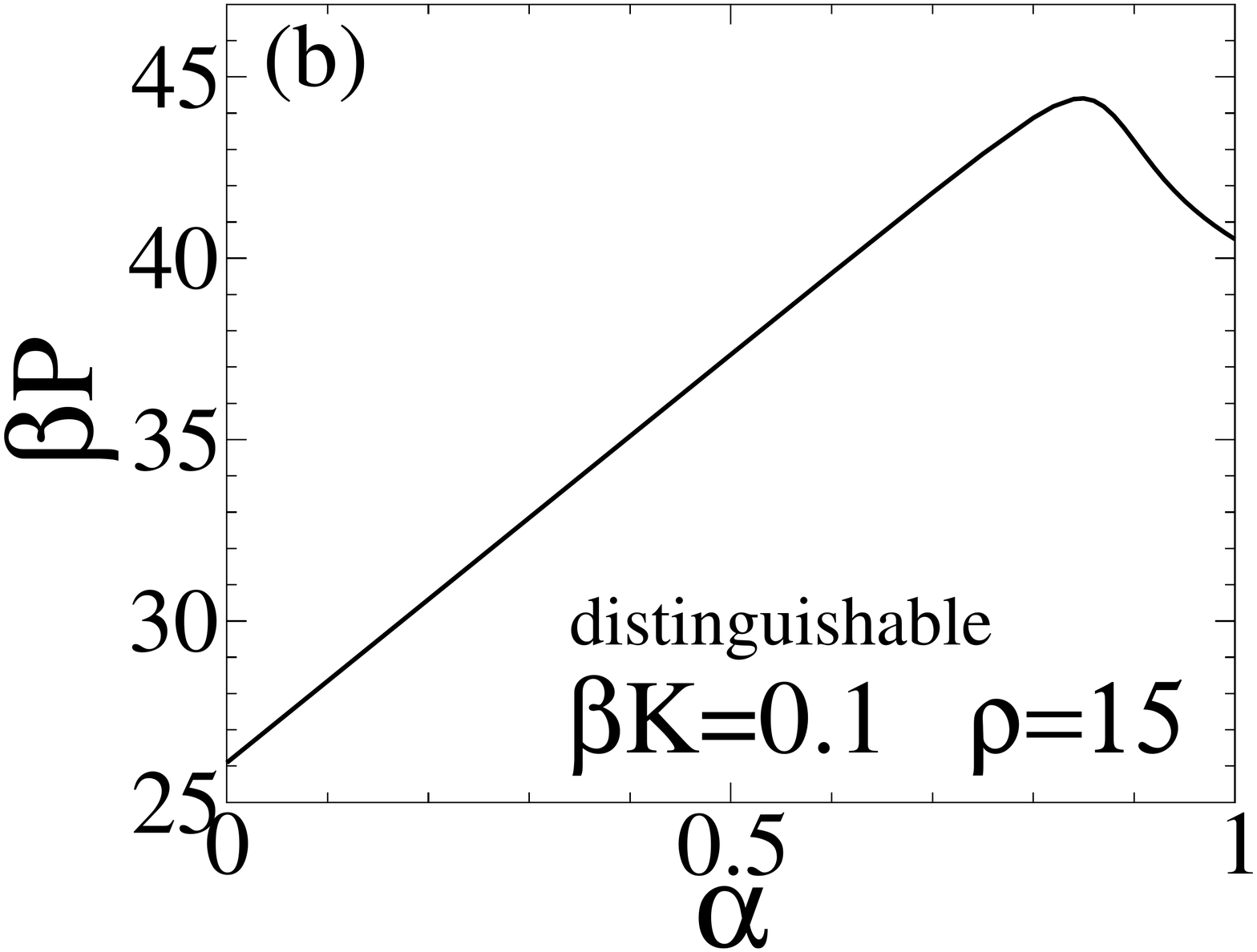}&
 \end{tabular}
 \end{center}
\caption{Pressure per lattice site as a function of $\alpha$.
The onset of a structural rearrangement into an alternating structure 
is accompanied by a reduced pressure.  }
\label{fig:pressure} 
\end{figure}

To illustrate how the structural rearrangement of adapting an alternating structure 
affects thermodynamic quantities of a system, 
in Fig. (\ref{fig:pressure}) we plot pressure per lattice site as a function of $\alpha$.   
Initially the pressure increases linearly with $\alpha$.  At the crossover, when the
distribution $p(n)$ becomes bimodal, this trend abruptly 
changes and pressure begins to decrease, reflecting the structural rearrangement
of a system that is concurrent with the reduction of internal tensions.    
\graphicspath{{figures/}}

So far we have examined strictly weak interactions, and all the results for $\beta K=0.1$, 
where the transformation into ordered alternating structure requires large densities.  

Additional feature of systems with strong interactions is that once an ordered 
alternative structure is attained, it is followed by additional regular transformations 
upon further squeezing of the system at around densities corresponding to integer
values, that is, when the occupation of a lattice site changes from $n\to n+1$ 
($n$ being an integer).  This results in a steplike structure of the pressure isotherm, 
which disappears at weak interactions.  Such a strong interaction case and the 
resulting regular transformations were carefully studied in \cite{Prestipino14,Prestipino15a}.
It was demonstrated by the researchers of the above references that the transformations
at $n\to n+1$ correspond to sharp crossovers rather than representing a true phase 
transitions.  

Limiting ourselves to weak
interactions in the present work, we eliminate the complication of the steplike 
behavior and demonstrate the onset of an ordered alternating structure through the 
continuous-like shape of $p(n)$.

\section{Two-component system}
\label{sec:model2}

In this section we consider a two-component 1d lattice model, where particles
of the same species repel, and those of the opposite species attract one another.  
The system Hamiltonian is 
\ba
H
&=& \frac{K}{2} \sum_{i=1}^L \bigg[n_i^+(n_i^+ - 1) + n_i^-(n_i^- - 1) - 2 n_i^+n_i^- \bigg] \nonumber\\ 
&+& \alpha K\sum_{i=1}^L (n_i^+-n_i^-) (n_{i+1}^+ - n_{i+1}^-), 
\label{eq:H12a}
\ea
where the two species are labeled as ``$+$'' and ``$-$'', in analogy to charged systems.  
The first line is for the 
interactions between particles occupying the same site (the second term on that line 
subtracts self-interaction introduced in the first term), and the second line is for particle 
interactions occupying neighboring sites.  We rewrite Eq. (\ref{eq:H12a}) as
\ba
H
&=&  \frac{K}{2}\sum_{i=1}^L (n_i^+ -n_i^-)^2  -  \frac{K}{2}\sum_{i=1}^L (n_i^+ + n_i^-) \nonumber\\ 
&+& \alpha K\sum_{i=1}^L (n_i^+-n_i^-) (n_{i+1}^+ - n_{i+1}^-), 
\label{eq:H12}
\ea
so that, apart for one term that in partition function is incorporated into the chemical potential, 
it is written in terms of "charge" per lattice site, $s_i=n_i^+-n_i^-$.

As for the one-component system, we consider both indistinguishable and distinguishable 
particles.  The partition function for the indistinguishable case is 
\be
\Xi_a = 
\sum_{n_1^+=0}^{\infty} \sum_{n_1^-=0}^{\infty}\dots \sum_{n_L^+=0}^{\infty} \sum_{n_L^-=0}^{\infty}
e^{-\beta H} \prod_{i=1}^L e^{\beta\mu (n_i^++n_i^-)}. 
\label{eq:Xi12a}
\ee
and that for the distinguishable one  
\be
\Xi_b = \sum_{n_1^+=0}^{\infty} \sum_{n_1^-=0}^{\infty}\dots \sum_{n_L^+=0}^{\infty} \sum_{n_L^-=0}^{\infty}
e^{-\beta H} \prod_{i=1}^L \frac{e^{\beta\mu (n_i^++n_i^-)}}{n_i^+!n_i^-!}.  
\label{eq:Xi12b}
\ee

Both partition functions can be transformed into the summations over $s_i=n_i^+-n_i^-$ (see Appendix \ref{sec:A4}),  
\be
\Xi_a = 
\!\!\!\! \sum_{s_1=-\infty}^{\infty} \!\! \dots  \!\! \sum_{s_L=-\infty}^{\infty}   
\prod_{i=1}^L e^{-\frac{\beta K}{2} s_i^2} e^{-\beta \alpha Ks_is_{i+1}}  \bigg(\frac{e^{\beta\mu' |s_i|} }{1-e^{2\beta\mu'}}\bigg)
\label{eq:Xia_12}
\ee
and
\be
\Xi_b = 
\!\!\!\! \sum_{s_1=-\infty}^{\infty} \!\! \dots  \!\! \sum_{s_L=-\infty}^{\infty}   
\prod_{i=1}^L e^{-\frac{\beta K}{2} s_i^2} e^{-\beta \alpha Ks_is_{i+1}}  
{\rm I}_{|s_i|} \big(2e^{\beta\mu'}\big), 
\label{eq:Xib_12}
\ee
where $\mu'=\mu + K/2$ and $I_n(x)$ in the second equation is the modified Bessel 
function of the first kind, 
\be
{\rm I}_{s}\big(2x\big) = x^{-s}\sum_{n=0}^{\infty}\frac{x^{2n}}{n!(n+s)!}.  
\ee
The corresponding transfer matrices are 
\be
T_a(s,s') =e^{-\frac{\beta K}{4} (s^2+s'^2)} e^{-\beta \alpha Ks s'} \bigg( \frac{e^{\frac{\beta\mu'}{2} (|s|+|s'|)} }{1-e^{2\beta\mu'}}\bigg), 
\label{eq:T12a}
\ee
and
\be
T_b(s,s') =e^{-\frac{\beta K}{4} (s^2+s'^2)} e^{-\beta \alpha Ks s'}   \sqrt{ {\rm I}_{|s|} \big[2e^{\beta\mu}\big] {\rm I}_{|s'|} \big[2e^{\beta\mu}\big]}.  
\label{eq:T12b}
\ee
Note that the matrices are similar to those in Eqs. (\ref{eq:Ta},\ref{eq:Tb}) for the one-component 
system.  What makes the two systems different is that the indices $s$ and $s'$ in 
Eqs. (\ref{eq:T12a},\ref{eq:T12b}) are for all integers, raising the possibility 
of the term $e^{-\beta \alpha Ks s'}$ 
to dominate the transfer matrix if $s$ and $s'$ have opposite sign and to eventual 
divergence of the partition function.  The divergence, however, can be switched off
for sufficiently low $\alpha$.  

For example, if we consider elements of the transfer matrix corresponding to $s'=-s$, 
\be
T_a(s,-s) = \frac{e^{-\frac{\beta K}{2} s^2(1-2\alpha)}e^{\beta\mu' |s|} }{1-e^{2\beta\mu'}},
\ee
we find that the possibility of divergence in the limit $s\to\infty$ is prevented if $\alpha\le 0.5$
(keeping in mind that $\mu'<0$).   For $\alpha>0.5$, no matter how negative $\mu'$, the
divergence can never be suppressed.  If we consider in turn the case $s>0$ and $s'=-1$,
\be
T_a(s,-1) = e^{-\frac{\beta K}{4}} e^{\frac{\beta\mu'}{2}}
\frac{e^{-\frac{\beta K}{4} s^2} e^{\beta \alpha Ks} e^{\frac{\beta\mu'}{2}s }}{1-e^{2\beta\mu'}},
\ee
we discover that the divergence in the limit $s\to\infty$ never arises for any $\alpha$, as the expression
is dominated by $e^{-\frac{\beta K}{4} s^2}$, which vanishes in the same limit.  
We conclude that divergent elements in the limit $s\to\infty$
are those that roughly satisfy $s'\approx -s $.  

The presence of the divergent terms implies that a charge, $\langle s\rangle$, at a single site 
becomes infinite, which, by the same token, implies that the occupation number at a single site 
diverges, pointing out to thermodynamic instability (also known as catastrophe).

\begin{figure}[h] 
 \begin{center}
 \begin{tabular}{rrrr}
  \includegraphics[height=0.16\textwidth,width=0.20\textwidth]{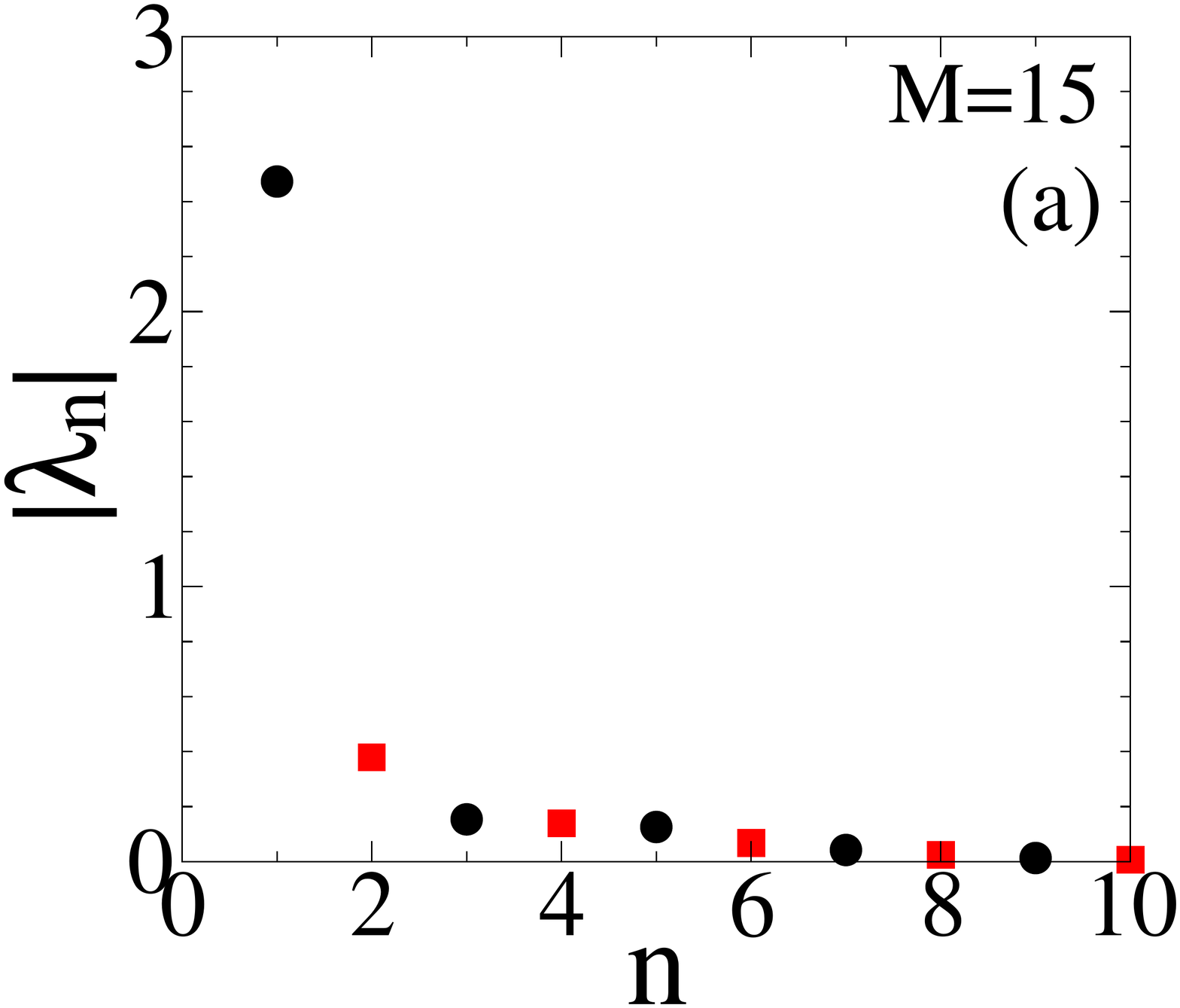}&
  \includegraphics[height=0.16\textwidth,width=0.20\textwidth]{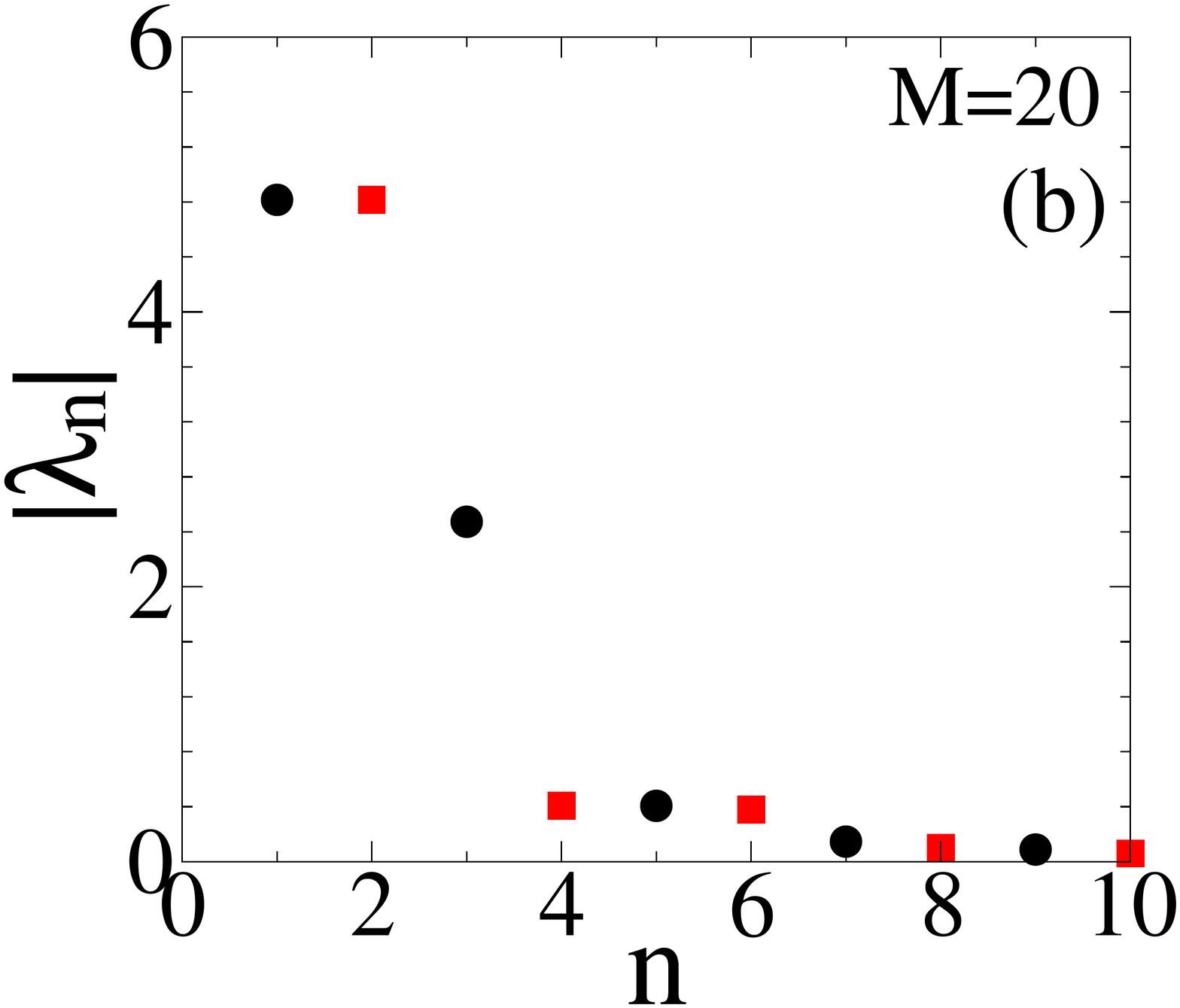}&
 \end{tabular}
 \end{center}
\caption{Ordered eigenvalues $|\lambda_1|>|\lambda_2|>\dots$ 
for $\beta K=0.1$, $\beta\mu'=-1$, and $\alpha=1$, for two different sizes of the 
transfer matrix $M\times M$.  }
\label{fig:lambda_A} 
\end{figure}

To see how the presence of the divergence plays its role, in Fig. (\ref{fig:lambda_A}) we plot 
$|\lambda_n|$ for $\alpha=1$ as a function of $n$, for different values of $M$, 
where $M$ is the size of a transfer matrix $M\times M$.  In the stable system, 
$|\lambda_n|$ converges in the limit $M\to\infty$.  In the unstable system, $|\lambda_n|$ diverge 
in the same limit.  In addition to blowing up of $|\lambda_n|$ as $M$ increases, we 
observe increased domination of the two initial eigenvalues, 
$|\lambda_1|\approx |\lambda_2|$, similar to what was seen for the 
one-component system in Fig. (\ref{fig:lambda}), and which indicates an ordered alternative 
structure.  For the two-component system, the ordering implies an alternating occupation 
of each site by a species "$+$" and "$-$".  
\graphicspath{{figures/}}

\begin{figure}[h] 
 \begin{center}
 \begin{tabular}{rrrr}
  \includegraphics[height=0.18\textwidth,width=0.22\textwidth]{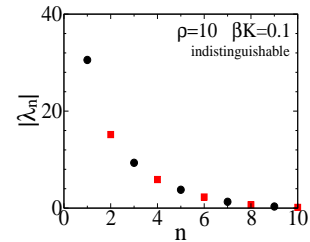}&
 \end{tabular}
 \end{center}
\caption{Ordered eigenvalues $|\lambda_1|>|\lambda_2|>\dots$.  
Compare with Fig. (\ref{fig:lambda}) for a one-component system.  Circles are for $n$-odd and 
squares for $n$-even. }
\label{fig:lambda12} 
\end{figure}

For comparison, in Fig. (\ref{fig:lambda12}) we plot $|\lambda_n|$ for a stable 
system at $\alpha=0.5$.  The results indicate that 
the two leading eigenvalues are separated even in the limit of high $\rho$, therefore,
never come to dominate.  The stability in this case implies the absence of an ordered 
alternating structure, and the region of stability corresponds to $\alpha\le 0.5$.
\graphicspath{{figures/}}

\begin{figure}[h] 
 \begin{center}
 \begin{tabular}{rrrr}
  \includegraphics[height=0.18\textwidth,width=0.22\textwidth]{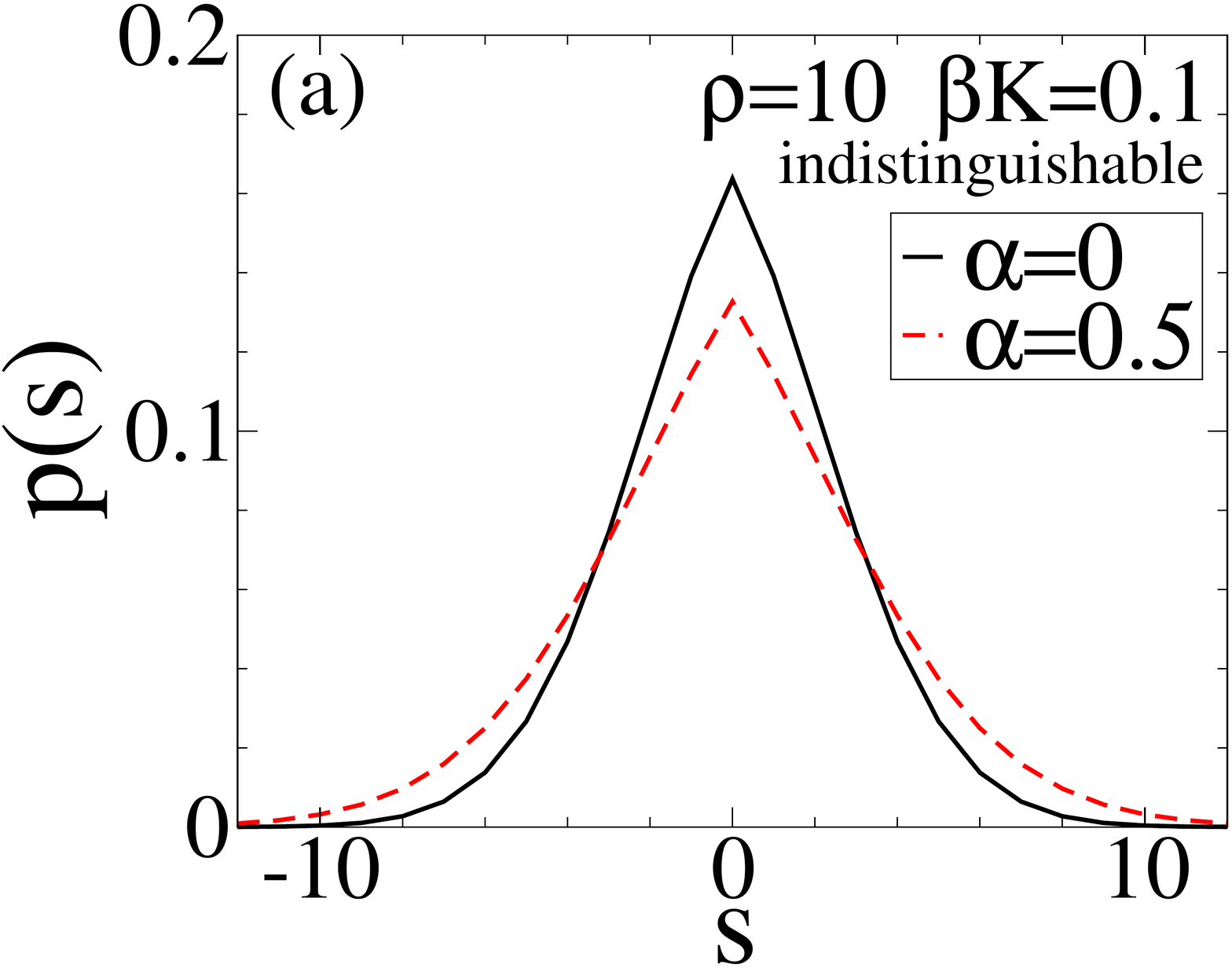}&
  \includegraphics[height=0.18\textwidth,width=0.22\textwidth]{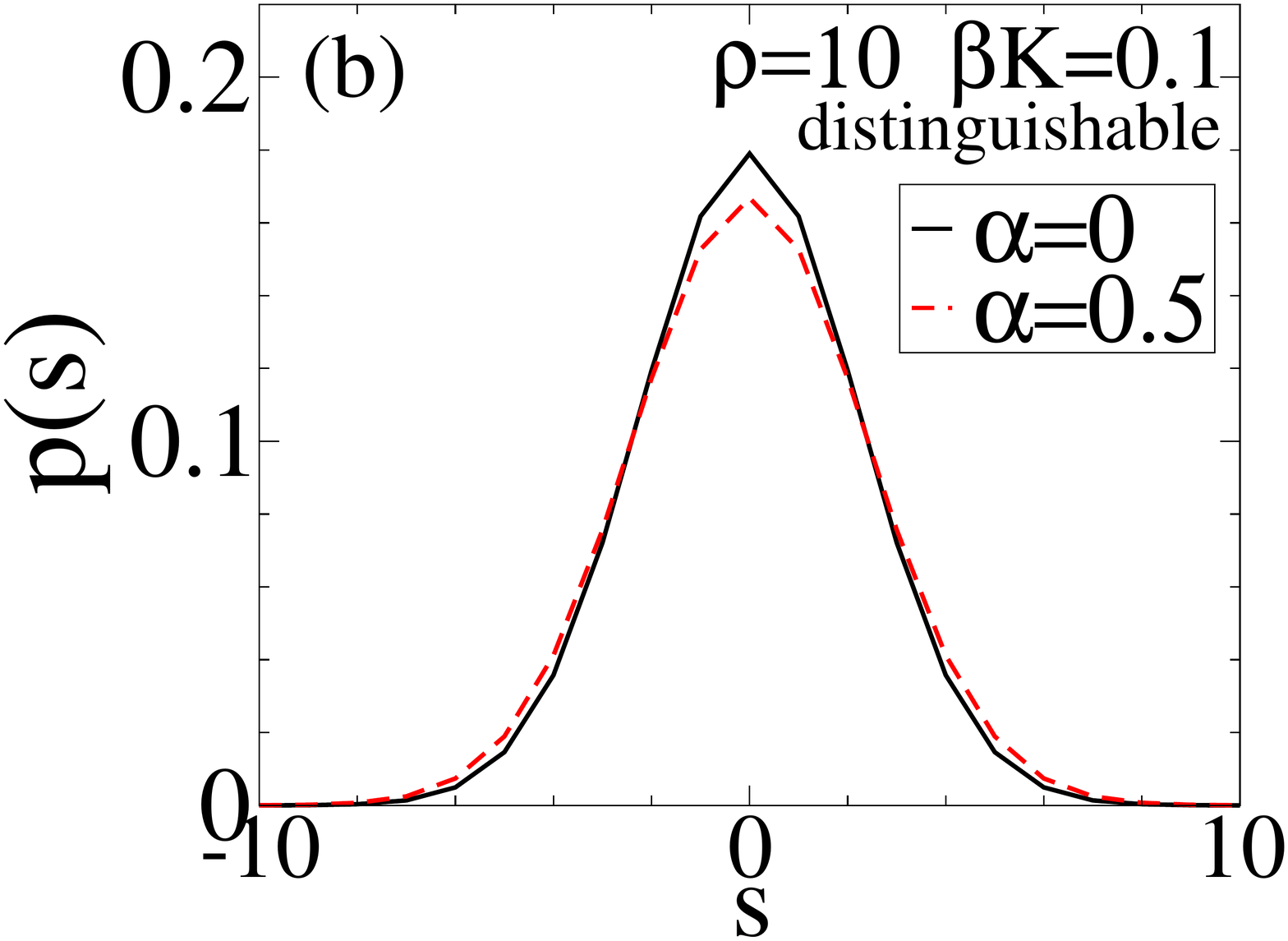}&
 \end{tabular}
 \end{center}
\caption{Distributions $p(s)$ for $\alpha=0$ and $\alpha=0.5$.  Apart for the broadening 
of $p(s)$ for $\alpha=0.5$, there is no evidence of a bimodal structure.  }
\label{fig:ps} 
\end{figure}

In search of possible structures in a stable system at $\alpha=0.5$, in 
Fig. (\ref{fig:ps}) we plot the distributions $p(s)=\phi_1^2(s)$ for $\alpha=0$ and $\alpha=0.5$.   
For $\alpha=0.5$ the distributions appear broader but otherwise fail to develop a bimodal structure.  
Indistinguishable particles show less response to the variation with $\alpha$, which we attribute 
to a higher entropy cost in adapting structured configuration.  
\graphicspath{{figures/}}

\begin{figure}[h] 
 \begin{center}
 \begin{tabular}{rrrr}
  \includegraphics[height=0.18\textwidth,width=0.22\textwidth]{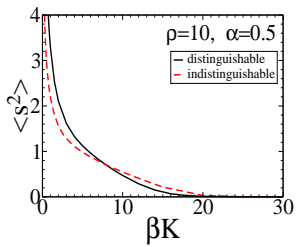}&
 \end{tabular}
 \end{center}
\caption{The fluctuations $\langle s^2\rangle$ as a function of $\beta K$ for $\alpha=0.5$ and $\rho=10$.  }
\label{fig:s2} 
\end{figure}

The most interesting structural feature of the two-component system at 
$\alpha=0.5$ is the formation of semi-stable pairs between particles of opposite species.  
In electrolytes these pairs are referred to as the Bjerrum pairs \cite{Yan93}.  
The presence of such pairs is evident in ``charge'' fluctuations, $\langle s^2\rangle$, plotted in 
Fig. (\ref{fig:s2}) as a function of $\beta K$ for $\rho=10$.  For large values of particle interactions
the fluctuations are suppressed, indicating that particles interchange sites not as free particles 
but as as permanent pairs.  
\graphicspath{{figures/}}

\subsection{significance of thermodynamic instability for real systems}

As stated above, the two-component lattice model is thermodynamically 
unstable for $\alpha>0.5$.  This criterion was derived from the grand canonical
ensemble.  To determine how this instability is manifested in real systems, 
we perform a sequence Monte Carlo simulations in a canonical ensemble for 
a two-component 1d lattice model for $\alpha>0.5$.   
The system is finite with periodic boundary conditions.  The system size is $L=1000$, there 
are $N^+=N^-=10000$ particles, and the interaction strength is $\beta K=0.1$.  
All simulations start with randomly distributed particles.

\begin{figure}[h] 
 \begin{center}
 \begin{tabular}{rrrr}
  \includegraphics[height=0.18\textwidth,width=0.23\textwidth]{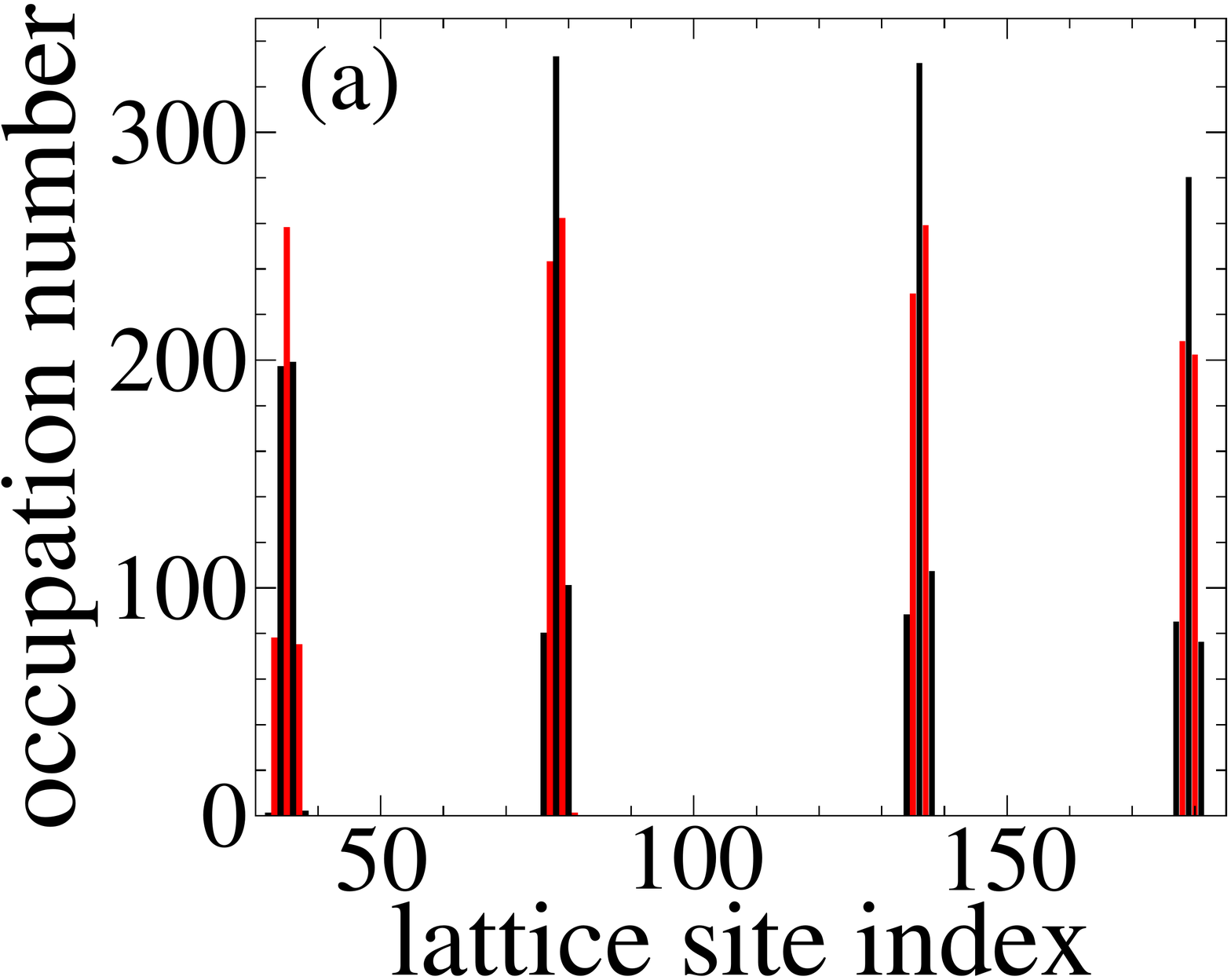}&
  \includegraphics[height=0.18\textwidth,width=0.22\textwidth]{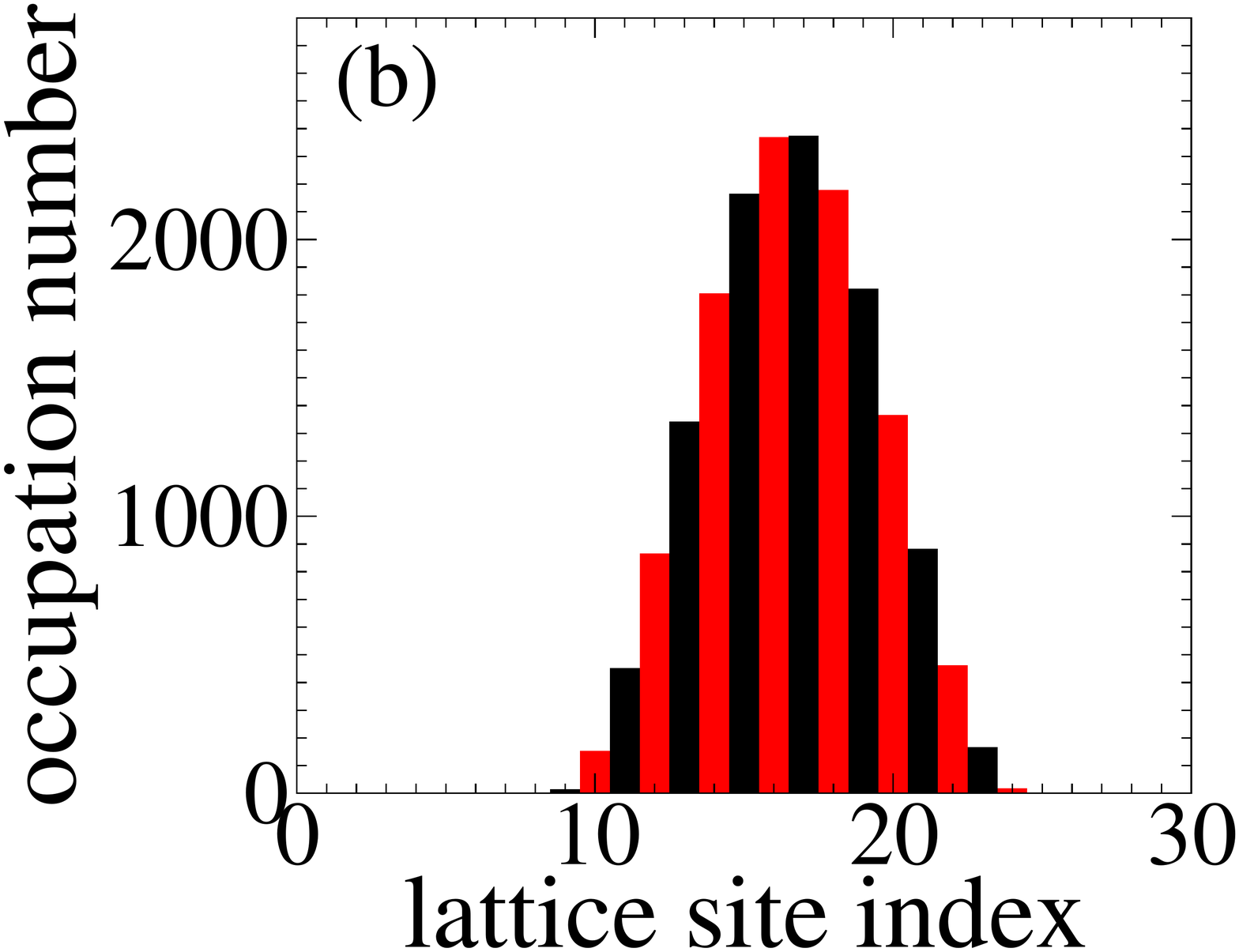}&
 \end{tabular}
 \end{center}
\caption{Monte Carlo simulation snapshots for $L=1000$ and
 $N^+=N^-=10000$ particles.  The interactions are set at $\beta K=0.1$ and the results are for 
 distinguishable particles.  Figure a) is for $\alpha=1$ and b) for $\alpha=0.55$.}  
\label{fig:cluster} 
\end{figure}

Configuration snapshots in Fig. (\ref{fig:cluster}) reveal that a system collapses into 
a finite number of clusters, the so called thermodynamic catastrophe \cite{Ruelle66a,Ruelle66b}.  
As all the sites are equivalent, the clusters form by spontaneous nucleation. 
For $\alpha=1$ in Fig. (\ref{fig:cluster}) (a) particles are distributed
over a several five-site clusters.  
As $\alpha$ decreases, larger clusters are preferred.  
A snapshot shown in Fig. (\ref{fig:cluster}) (b) for $\alpha=0.55$ indicates that 
an entire system exists as a single sixteen-site cluster.  The cluster disintegrates 
for $\alpha\le 0.54$, as there are not enough particles to reorganize into  
a larger cluster.  This does not imply thermodynamic stability, however.  Changing 
the system size would produce another collapse into larger clusters.  
\graphicspath{{figures/}}

To understand the dependence of the cluster size on $\alpha$, we consider a number of 
clusters.  The smallest possible cluster is comprised of two sites.  The energy
of this cluster is obtained from the Hamiltonian in Eq. (\ref{eq:H12}).  Assuming that two sites have 
the same occupation number $n/2$, where $n$ is the total number of particles in the cluster,
the energy is
\be
\frac{E^{(2)}}{K} = \frac{n}{2}\bigg(\frac{n}{2}-1\bigg) - \alpha \bigg(\frac{n}{2}\bigg)^2 = 
-\bigg(\frac{n}{2}\bigg)^2(\alpha-1) - \frac{n}{2}, 
\ee
where the first term, which is proportional to $n^2$, is positive if $\alpha<1$, therefore, 
the system does not collapse into a single cluster for $\alpha\le1$, as the cluster energy
eventually becomes positive for large $n$.  

Next we consider the three-site cluster. If the total number of particles in 
the cluster is $n$, and the occupation of the central site is $n_0$ and that of each flanking 
site is $n-n_0$ (we assume symmetricity of a cluster), then based on Eq. (\ref{eq:H12}) the 
energy is 
\be
\frac{E^{(3)}}{K} = \frac{1}{2}n_0^2 + \frac{1}{4}(n-n_0)^2 - \alpha n_0(n-n_0) - \frac{n}{2}, 
\ee
This energy, optimized with respect to $n_0$, becomes
\be
\frac{E_0^{(3)}}{K} =  -\frac{n^2}{2}\bigg(\frac{2\alpha^2-1}{4\alpha+3}\bigg)  - \frac{n}{2}.  
\ee
At $\alpha=1/\sqrt{2}\approx 0.707$ the term proportional to $n^2$ becomes
positive and the three-site cluster disintegrates as its energy becomes positive for large $n$.  

We can repeat the same calculations for a four-site cluster, whose optimized energy is (optimized
clusters are symmetric)
\be
\frac{E_0^{(4)}}{K} =  -\frac{n^2}{4}\bigg(\frac{\alpha^2+\alpha-1}{\alpha+2}\bigg) - \frac{n}{2}, 
\ee
and the cluster disintegrates at $\alpha=(\sqrt{5}-1)/2\approx 0.618$, as the first term becomes positive.  
Each consecutive cluster disintegrating at $\alpha$ closer to $0.5$.  For example, 
a five-site cluster disintegrates at $\alpha=1/\sqrt{3}\approx 0.577$.

\begin{figure}[h] 
 \begin{center}
 \begin{tabular}{rrc}
  \includegraphics[height=0.18\textwidth,width=0.23\textwidth]{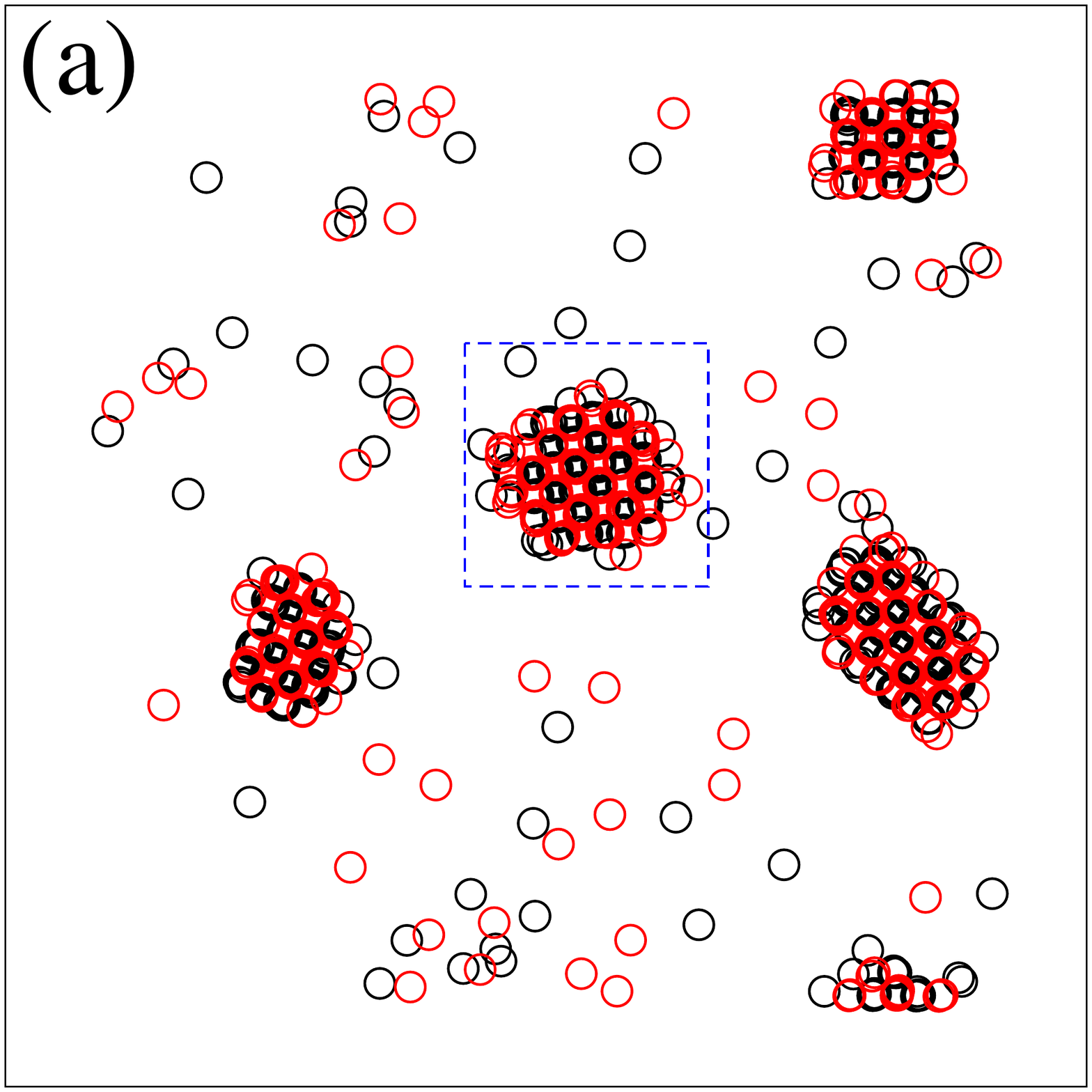}&
  \includegraphics[height=0.18\textwidth,width=0.23\textwidth]{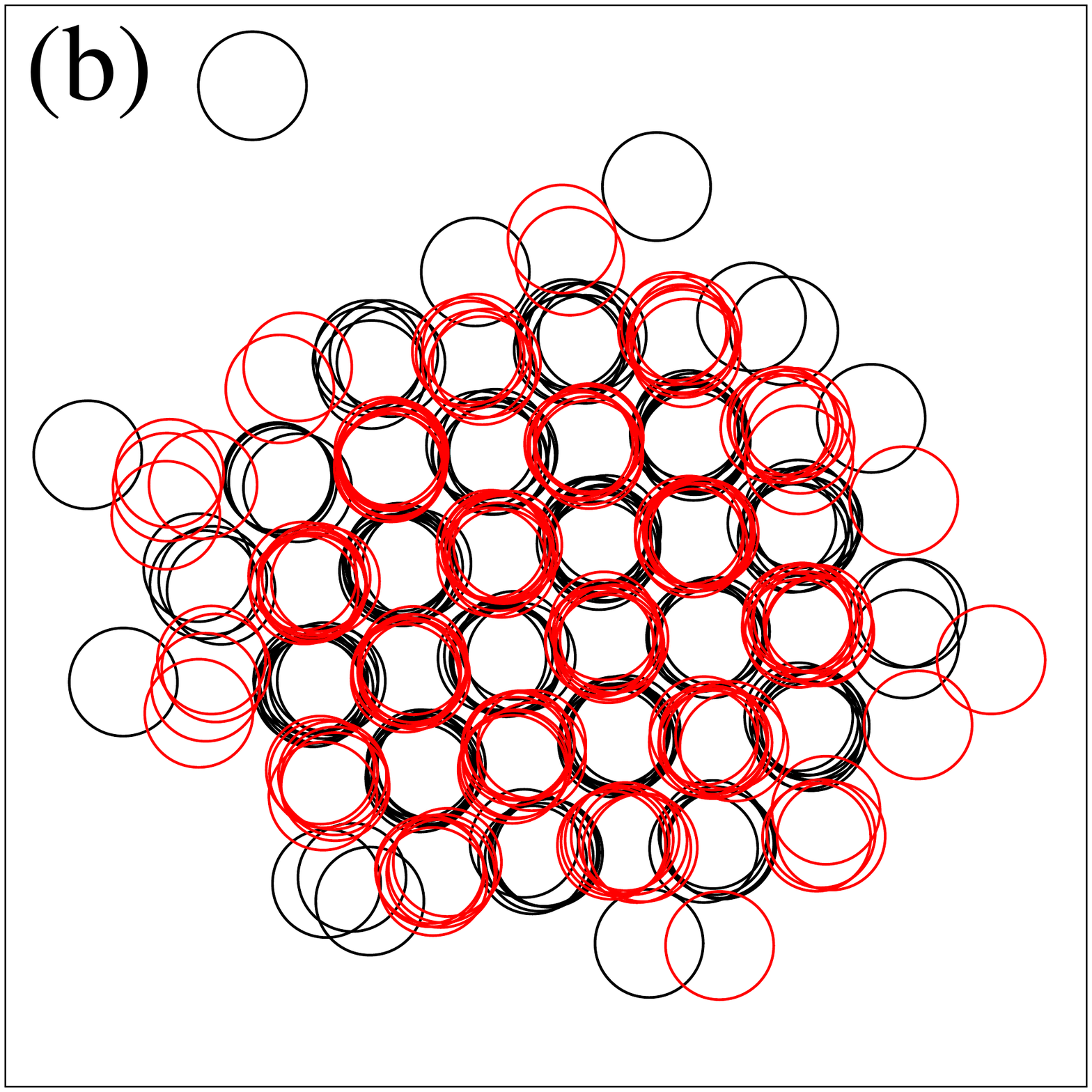}\\
 \end{tabular}
 \end{center}
\caption{Configuration snapshots of a two-component penetrable-sphere system in 2d.  
The red and black circles indicate particles of different species.  Particles do not interact
unless they overlap.  At an overlap, the interactions are $\beta u(r<\sigma)=\pm1$.  
The first figure contains $N=1000$ particles and the second one $N=240$.  } 
\label{fig:solid}
\end{figure}

The observed thermodynamic catastrophe is not an artifact of a lattice 
model, and a similar behavior has been observed for a two-component penetrable-spheres 
\cite{Frydel17}.  A snapshot 
for a 2d two-component Penetrable-Sphere model is shown in Fig. (\ref{fig:solid}), which reveals 
a similar catastrophe, characterized by a system collapse into small number of large clusters.  
\graphicspath{{figures/}}
On the other hand, thermodynamic catastrophe does not occur for a 
two-component Gaussian core model \cite{Frydel18}, indicating that this
system does not fulfill the conditions of instability.

In the one-component system the criterion $\alpha> 0.5$ tells us that
a system under certain conditions of density and interaction strength 
can adopt an ordered alternating structure.  The condition $\alpha> 0.5$, 
therefore, is necessary but not sufficient for this situation.    
In the two-component system, on the other hand, the criterion $\alpha> 0.5$ 
tells us that a system is thermodynamically unstable under any conditions,
no matter what its density and the interaction strength, as long as a system 
is in thermodynamic limit.   The criterion $\alpha> 0.5$ for the two-component
system, therefore, is necessary and sufficient.

\subsection{criterion for thermodynamic catastrophe}

Thermodynamic catastrophe of a one-component system was first investigated by Ruelle 
and Fisher in \cite{Ruelle66a,Ruelle66b}.  The condition for thermodynamic instability 
for these systems is $\int d{\bf r}\, u(r) < 0$, or using the Fourier transformed pair 
potential $\tilde u(k)$, the same condition is stated as $\tilde u(0)<0$.  
This implies that the potential needs to have an attractive part and
a non-divergent (soft) core.  An example of such a potential is a double Gaussian potential 
investigated in \cite{Prestipino15b,Prestipino16}.  

Based on our results for a lattice model and its connection to real penetrable particles, 
we conclude that the condition for thermodynamic instability for a two-component system 
are provided by the LLWL criterion.  That is, if $\tilde u(k)$ is negative for some value of $k$, 
then the two-component system is thermodynamically unstable.  In turn, if $\tilde u(k)$ is 
positive everywhere, then the system is stable.  This justifies why thermodynamic catastrophe 
is observed for the two-component Penetrable-Sphere but not the Gaussian core model.  

\section{Conclusion}
\label{sec:conclusion}

The present work investigates a 1d lattice model with multiple occupations as a simple
representation of penetrable particles.  Starting with non-interacting particles, we
distinguish between different ways of counting configurations, by treating particles 
as either indistinguishable or distinguishable, leading to two different partition functions.  
The indistinguishable case is representative of growth models, and the distinguishable
case is representative of liquids.  

For a one-component case we discover two classes of behavior, depending on weather 
$\alpha>0.5$ or $\alpha\le 0.5$.  For $\alpha>0.5$, under the conditions of large $\rho$ 
and/or large $\beta K$, systems form an ordered and alternating structure of occupied 
versus empty sites.  For $\alpha\le 0.5$ such structural reorganization never take place.  
Because the condition $\alpha<0.5$ does not guarantee the presence of an alternating
structure, we say that this condition is necessary but not sufficient.  

For the two-component system, where particles of the same species repel and of opposite
species attract each other, we find that systems with $\alpha>0.5$ are thermodynamically
unstable, and such systems exhibit thermodynamic catastrophe \cite{Ruelle66a,Ruelle66b}.
In this case, the criterion $\alpha>0.5$ is necessary and sufficient for thermodynamic
instability, as long as the system is in thermodynamic limit.  Even a dilute system with 
weak interactions eventually collapses.  

Thermodynamic instability was observed in the Penetrable-Sphere model, but not in the Gaussian 
core model.  Consequently, we conclude that the LLWL criterion devised for a one-component 
system apply to a two-component system for predicting the conditions of thermodynamic instability.

A shortcoming of a 1d lattice model is that it does not undergo a true phase transition.  A more realistic  
representation of penetrable particles would require working in 2d or higher dimension, where 
such transition becomes feasible.  Some aspects of a lattice model in higher dimension 
studied in \cite{Hansen04} for repulsive interactions. In future we plan to investigate a 
two-component lattice model
in 2d or a higher dimension, in order to study a gas-liquid phase transition and the role of
Bjerrum pairs in the transition mechanism \cite{Frydel18}.

\appendix 
\section{Derivation of $p(n)$ within the transfer matrix method}
\label{sec:A1}

In Sec. (\ref{sec:model1}) in Eq. (\ref{eq:p_n0}) we provide the relation $p(n)=\phi_1^2(n)$  
of the transfer matrix method, for the probability $p(n)$ that a single lattice site is 
occupied by $n$ particles.  The relation can be rigorously derived from the formal definition 
\be
p(n) = \frac{1}{\Xi} \sum_{n_2=0}^{\infty}\dots\sum_{n_L=0}^{\infty} T(n,n_2) T(n_2,n_3)\dots T(n_L,n),
\label{eq:A0}
\ee
although the derivational steps are omitted.  The aim of this appendix is to provide the omitted 
derivation.  Note that $p(n)$ is normalized by construction, $\sum_{n=0}^{\infty}p(n)=1$.  

The first step is to rewrite the definition in Eq. (\ref{eq:A0}) in shorthanded form, 
using the matrix algebra nomenclature that is more convenient for carrying out subsequent 
matrix operations, 
\be
p(n) = \frac{1}{\Xi} ({\bf T}^L)(n,n), 
\label{eq:A1}
\ee
where $({\bf T}^L)(n,n')$ designates the $(n,n')$ element of the matrix ${\bf T}^L$, where 
${\bf T}^L$ is the product matrix generated by multiplying ${\bf T}$ by itself $L$-times.
In the next step we apply the eigendecomposition
\be
T(n,n') = \sum_{k=1}^{\infty} \lambda_k  Q(n,k) Q^{-1}(k,n'), 
\ee
where 
${\bf Q}$ is the square matrix whose $k$-th column is the eigenvector $\phi_k(n)$ of 
the transfer matrix ${\bf T}$, ${\bf Q}^{-1}$ is the inverse of ${\bf Q}$ such that 
\be
{\bf Q}{\bf Q}^{-1} = {\bf I},
\ee
where ${\bf I}$ is the identity matrix.  Eigendecomposition applied to 
${\bf T}^L$ yields 
\be
({\bf T}^L)(n,n') = \sum_{k=1}^{\infty} \lambda_{k}^L Q(n,k) Q^{-1}(k,n').  
\ee
The probability $p(n)$ in Eq. (\ref{eq:A1}) can now be written as
\be
p(n) = \frac{1}{\Xi}\sum_{k=1}^{\infty} \lambda_{k}^L Q(n,k)Q^{-1}(k,n).  
\ee
Because the transfer matrix is real and symmetric, ${\bf Q}^{-1}={\bf Q}^{T}$, 
where ${\bf Q}^{T}$ is the transpose of ${\bf Q}$, and we write 
\be
p(n) = \frac{\sum_{k=1}^{\infty} \lambda_{k}^L Q^2(n,k)}{\sum_{k=1}^{\infty}\lambda_k^L},
\ee
where we used $\Xi={\rm Tr} \, {\bf T}^L=\sum_{k=1}^{\infty}\lambda_k^L$.  
Since the columns of the matrix ${\bf Q}$ correspond to eigenvectors 
$\phi_k$, we get 
\be
p(n) = \frac{\sum_{k=1}^{\infty} \lambda_{k}^L \phi_k^2(n)}{\sum_{k=1}^{\infty}\lambda_k^L}. 
\ee
In the final step we take the thermodynamic limit, $L\to\infty$, in which the above expression 
reduces to  
\be
p(n) = \phi_1^2(n),
\label{eq:A2}
\ee
which recovers the result of Eq. (\ref{eq:p_n0}).

\section{Derivation of $p_m(n,n')$ within the transfer matrix method}
\label{sec:A2}

In this appendix we derive the expression for the probability $p_m(n,n')$, that a number 
of particles at two lattice sites separated by $m$ sites is $n$ and $n'$.  The derived result 
appears in Eq. (\ref{eq:p12}) but no derivation is provided.  We start with the formal definition
for $p_m(n,n')$, 
\ba
p_m(n,n') &=& \frac{1}{\Xi} \sum_{n_2=0}^{\infty}\dots\sum_{n_{m}=0}^{\infty}T(n,n_2) T(n_2,n_3)\dots T(n_m,n') \nonumber\\
&&\dots \sum_{n_{m+2}=0}^{\infty}\dots \sum_{n_L=0}^{\infty} T(n',n_{m+2})\dots T(n_L,n). \nonumber\\ 
\ea
Note that $p_m(n,n')$ is normalized by construction, $\sum_{n=0}^{\infty}\sum_{n'=0}^{\infty}p_m(n,n')=1$.
Using matrix algebra nomenclature, the above expression can be shorthanded into 
\be
p_m(n,n') = \frac{1}{\lambda_1^L} ({\bf T}^m)(n,n') ({\bf T}^{L-m})(n',n).  
\ee
Then eigendecomposition yields  
\be
({\bf T}^m)(n,n') = \sum_{k} \lambda_{k}^m Q(n,k) Q^{-1}(k,n'), 
\ee
and
\be
({\bf T}^{L-m})(n',n) = \sum_{k} \lambda_{k}^{L-m} Q(n',k) Q^{-1}(k,n),
\ee
leading to 
\ba
p_m(n,n') &=& \frac{1}{\Xi} \bigg(\sum_{k} \lambda_{k}^{m} Q(n,k) Q^{-1}(k,n')\bigg)\nonumber\\
&\times&\bigg(\sum_{k'} \lambda_{k'}^{L-m} Q(n',k') Q^{-1}(k',n)\bigg).  
\ea
Further simplifications follow from the fact that for a real and symmetric matrix ${\bf T}$, 
${\bf Q}^{-1}={\bf Q}^{T}$, and the columns of the matrix ${\bf Q}$ correspond to 
eigenvectors $\phi_k$.  This leads to   
\be
p_m(n,n') = \frac{\sum_{k=1}^{\infty}\sum_{k'=1}^{\infty}
\lambda_{k}^{m}\lambda_{k'}^{L-m}  \phi_k(n)\phi_{k'}(n)  \phi_k(n') \phi_{k'}(n')}
{\sum_{k=1}^{\infty}\lambda_k^L},
\ee 
where we used $\Xi=\sum_{k=1}^{\infty}\lambda_k^L$.
The final reduction comes from taking the thermodynamic limit, $L\to\infty$, in which case 
only $k'=1$ does not vanish, leading to the final result
\be
p_m(n,n') = \sum_{k=1} ^{\infty}
\bigg(\frac{\lambda_k}{\lambda_1}\bigg)^m  \phi_k(n) \phi_k(n') \phi_{1}(n) \phi_{1}(n').  
\ee 
Using the relation $p(n)=\phi_1^2(n)$ of the previous section we get 
\be
\frac{p_m(n,n')}{p(n)p(n')} 
= 1 + \sum_{k=2}^{\infty} \bigg(\frac{\lambda_k}{\lambda_1}\bigg)^m  \frac{\phi_k(n) \phi_k(n')}{\phi_{1}(n) \phi_{1}(n')},
\ee 
which agrees with Eq. (\ref{eq:p12}).

\section{Reduction of the partition functions for the two-component system}
\label{sec:A4}
We reduce the partition functions of the two-component systems, from the 
$2L$-summation to the $L$-summation, by considering the simple case $L=3$,
however, the procedure is general and valid for any $L$.  For distinguishable 
particles the partition function for three lattice sites, obtained from Eq. (\ref{eq:Xi12a}), is
\ba
\Xi_a &=& 
\sum_{n_1^+=0}^{\infty} \sum_{n_1^-=0}^{\infty} \sum_{n_2^+=0}^{\infty} \sum_{n_2^-=0}^{\infty} 
\sum_{n_3^+=0}^{\infty} \sum_{n_3^-=0}^{\infty} \nonumber\\ &&
e^{-\frac{\beta K}{2}(n_1^+-n_1^-)^2 } e^{-\alpha \beta K(n_1^+-n_1^-)(n_{2}^+-n_{2}^-) } e^{\beta\mu' (n_1^++n_1^-)}\nonumber\\
&\times&
e^{-\frac{\beta K}{2}(n_2^+-n_2^-)^2 } e^{-\alpha \beta K(n_2^+-n_2^-)(n_{3}^+-n_{3}^-) } e^{\beta\mu' (n_2^++n_2^-)}\nonumber\\
&\times&
e^{-\frac{\beta K}{2}(n_3^+-n_3^-)^2 } e^{-\alpha \beta K(n_3^+-n_3^-)(n_{1}^+-n_{1}^-) } e^{\beta\mu' (n_3^++n_3^-)}
\nonumber\\
\label{eq:D0}
\ea
where for the sake of clarity we write down every term explicitly.  We also recall that $\mu'=\mu+K/2$.  
The above expression can be shortened by using $s_i=n_i^+-n_i^-$, 
\ba
\Xi_a &=& 
\sum_{n_1^+=0}^{\infty} \sum_{n_1^-=0}^{\infty} \sum_{n_2^+=0}^{\infty} \sum_{n_2^-=0}^{\infty} 
\sum_{n_3^+=0}^{\infty} \sum_{n_3^-=0}^{\infty} \nonumber\\ &&
e^{-\frac{\beta K}{2}s_1^2 } e^{-\alpha \beta Ks_1s_2} e^{\beta\mu' (n_1^++n_1^-)}\nonumber\\
&\times&
e^{-\frac{\beta K}{2}s_2^2 } e^{-\alpha \beta Ks_2s_3} e^{\beta\mu' (n_2^++n_2^-)}\nonumber\\
&\times&
e^{-\frac{\beta K}{2}s_3^2 } e^{-\alpha \beta Ks_3s_1} e^{\beta\mu' (n_3^++n_3^-)},
\nonumber
\ea
which by itself does not yet transform the summation.  In order to transform the above result into 
summation in terms of $s_i$ we have to carry out partial summations of some of the terms, leading to
\ba
&&\Xi_a = \sum_{s_1=-\infty}^{\infty} \sum_{s_2=-\infty}^{\infty} \sum_{s_3=-\infty}^{\infty} \nonumber\\ &&
e^{-\frac{\beta K}{2}s_1^2 } e^{-\alpha \beta Ks_1s_2} 
\bigg(\sum_{n_1^+=0}^{\infty}\sum_{n_1^-=0}^{\infty} e^{\beta\mu' (n_1^++n_1^-)} \delta_{s_1,n_1^+-n_1^-}\bigg)
\nonumber\\
&\times&
e^{-\frac{\beta K}{2}s_2^2 } e^{-\alpha \beta Ks_2s_3} 
\bigg(\sum_{n_2^+=0}^{\infty}\sum_{n_2^-=0}^{\infty} e^{\beta\mu' (n_2^++n_2^-)} \delta_{s_2,n_2^+-n_2^-}\bigg)
\nonumber\\
&\times&
e^{-\frac{\beta K}{2}s_3^2 } e^{-\alpha \beta Ks_3s_1} 
\bigg(\sum_{n_3^+=0}^{\infty}\sum_{n_3^-=0}^{\infty} e^{\beta\mu' (n_3^++n_3^-)} \delta_{s_3,n_3^+-n_3^-}\bigg),
\nonumber\\
\ea
where $\delta_{ij}$ is the Kronecker delta function.  To complete the transformation we need to calculate
\be
f_a(s_i)=\bigg(\sum_{n_i^+=0}^{\infty}\sum_{n_i^-=0}^{\infty} e^{\beta\mu' (n_i^++n_i^-)} \delta_{s_i,n_i^+-n_i^-}\bigg), 
\label{eq:D1}
\ee
where the solution is found to be
\be
f_a(s_i)
= \frac{e^{\beta\mu'|s_i|}}{1-e^{2\beta\mu'}}, 
\label{eq:D2}
\ee
and where to obtain it we summed up all the terms corresponding to a given $s_i$.  
For example, for $s_i=0,1,2$ using Eq. (\ref{eq:D1}) we get 
\be
f_a(0) = \sum_{n=0}^{\infty} e^{2\beta\mu' n} = \frac{1}{1-e^{2\beta\mu'}}, 
\ee
\be
f_a(1) = e^{\beta\mu'}\sum_{n=0}^{\infty} e^{2\beta\mu' n} = \frac{e^{\beta\mu'}}{1-e^{2\beta\mu'}},
\ee
\be
f_a(2) = e^{2\beta\mu'}\sum_{n=0}^{\infty} e^{2\beta\mu' n} = \frac{e^{2\beta\mu'}}{1-e^{2\beta\mu'}}.
\ee
Furthermore, it turns out that $f_a(1)=f_a(-1)$, $f_a(2)=f_a(-2)$, etc., confirming the validity of the 
expression in Eq. (\ref{eq:D2}).  Consequently, the transformed partition function of Eq. (\ref{eq:D0}) 
becomes 
\ba
\Xi_a &=& \sum_{s_1=-\infty}^{\infty} \sum_{s_2=-\infty}^{\infty} \sum_{s_3=-\infty}^{\infty} \nonumber\\ &&
e^{-\frac{\beta K}{2}s_1^2 } e^{-\alpha \beta Ks_1s_2} \frac{e^{\beta\mu'|s_1|}}{1-e^{2\beta\mu'}}\nonumber\\
&\times&
e^{-\frac{\beta K}{2}s_2^2 } e^{-\alpha \beta Ks_2s_3} \frac{e^{\beta\mu'|s_2|}}{1-e^{2\beta\mu'}}\nonumber\\
&\times&
e^{-\frac{\beta K}{2}s_3^2 } e^{-\alpha \beta Ks_3s_1} \frac{e^{\beta\mu'|s_3|}}{1-e^{2\beta\mu'}}.
\nonumber\\
\ea
The procedure applies to any number of lattice sites.  

We apply a similar procedure for indistinguishable particles.  For the case $L=3$,
the partition function obtained from Eq. (\ref{eq:Xi12b}) is 
\ba
\Xi_b &=& 
\sum_{n_1^+=0}^{\infty} \sum_{n_1^-=0}^{\infty} \sum_{n_2^+=0}^{\infty} \sum_{n_2^-=0}^{\infty} 
\sum_{n_3^+=0}^{\infty} \sum_{n_3^-=0}^{\infty} \nonumber\\ &&
e^{-\frac{\beta K}{2}(n_1^+-n_1^-)^2 } e^{-\alpha \beta K(n_1^+-n_1^-)(n_{2}^+-n_{2}^-) } 
\frac{e^{\beta\mu' (n_1^++n_1^-)}}{n_1^+!n_1^-!}\nonumber\\&\times&
e^{-\frac{\beta K}{2}(n_2^+-n_2^-)^2 } e^{-\alpha \beta K(n_2^+-n_2^-)(n_{3}^+-n_{3}^-) } 
\frac{e^{\beta\mu' (n_2^++n_2^-)}}{n_2^+!n_2^-!} \nonumber\\&\times&
e^{-\frac{\beta K}{2}(n_3^+-n_3^-)^2 } e^{-\alpha \beta K(n_3^+-n_3^-)(n_{1}^+-n_{1}^-) } 
\frac{e^{\beta\mu' (n_3^++n_3^-)}}{n_3^+!n_3^-!} \nonumber\\
\label{eq:D2}
\ea
then transforming it into the summations over $s_i$ we get 
\ba
\Xi_b &=& 
\sum_{s_1=-\infty}^{\infty} \sum_{s_2=-\infty}^{\infty} \sum_{s_3=-\infty}^{\infty} \nonumber\\ &&
e^{-\frac{\beta K}{2}s_1^2 } e^{-\alpha \beta Ks_1s_2 } 
\bigg(\sum_{n_1^+=0}^{\infty}\sum_{n_1^-=0}^{\infty} \frac{e^{\beta\mu' (n_1^++n_1^-)}}{n_1^+!n_1^-!} \delta_{s_1,n_1^+-n_1^-}\bigg)
\nonumber\\&\times&
e^{-\frac{\beta K}{2}s_2^2 } e^{-\alpha \beta Ks_2s_3 } 
\bigg(\sum_{n_2^+=0}^{\infty}\sum_{n_2^-=0}^{\infty} \frac{e^{\beta\mu' (n_2^++n_2^-)}}{n_2^+!n_2^-!} \delta_{s_2,n_2^+-n_2^-}\bigg)
\nonumber\\&\times&
e^{-\frac{\beta K}{2}s_3^2 } e^{-\alpha \beta Ks_3s_1 } 
\bigg(\sum_{n_3^+=0}^{\infty}\sum_{n_3^-=0}^{\infty} \frac{e^{\beta\mu' (n_3^++n_3^-)}}{n_3^+!n_3^-!} \delta_{s_3,n_3^+-n_3^-}\bigg),
\nonumber\\
\label{eq:D2}
\ea
where it now remains to obtain 
\be
f_b(s_i) = 
\bigg(\sum_{n_i^+=0}^{\infty}\sum_{n_i^-=0}^{\infty} \frac{e^{\beta\mu' (n_i^++n_i^-)}}{n_i^+!n_i^-!} \delta_{s_i,n_i^+-n_i^-}\bigg).  
\ee
The result turns out to be
\be
f_b(s_i) = {\rm I}_{|s_i|} \big(2e^{\beta\mu'}\big),
\label{eq:D3}
\ee
where 
\be
{\rm I}_{s}\big(2x\big) = x^{-s}\sum_{n=0}^{\infty}\frac{x^{2n}}{n!(n+s)!},  
\ee
is the modified Bessel function of the first kind.  To validate this expression we proceed 
as before.  The results for $s_i=0,1,2$ are
\be
f_b(0) = \sum_{n=0}^{\infty}\frac{e^{2\beta\mu'n}}{n!n!} = {\rm I}_{0}\big(2e^{\beta\mu'}\big), 
\ee
\be
f_b(1) = e^{\beta\mu'}\sum_{n=0}^{\infty}\frac{e^{2\beta \mu'n}}{n!(n+1)!} = {\rm I}_{1}\big(2e^{\beta\mu'}\big), 
\ee
\be
f_b(2) = e^{2\beta\mu'}\sum_{n=0}^{\infty}\frac{e^{2\beta\mu'n}}{n!(n+1)!} = {\rm I}_{2}\big(2e^{\beta\mu'}\big), 
\ee
then $f_b(1)=f_b(-1)$, $f_b(2)=f_b(-2)$, etc., confirming the result in Eq. (\ref{eq:D3}).  
The transformed partition function becomes 
\ba
\Xi_b &=& 
\sum_{s_1=-\infty}^{\infty} \sum_{s_2=-\infty}^{\infty} \sum_{s_3=-\infty}^{\infty} \nonumber\\ &&
e^{-\frac{\beta K}{2}s_1^2 } e^{-\alpha \beta Ks_1s_2 } {\rm I}_{|s_1|} \big(2e^{\beta\mu'}\big)
\nonumber\\&\times&
e^{-\frac{\beta K}{2}s_2^2 } e^{-\alpha \beta Ks_2s_3 } {\rm I}_{|s_2|} \big(2e^{\beta\mu'}\big)
\nonumber\\&\times&
e^{-\frac{\beta K}{2}s_3^2 } e^{-\alpha \beta Ks_3s_1 } {\rm I}_{|s_3|} \big(2e^{\beta\mu'}\big). 
\nonumber\\
\ea
Again, the procedure is valid for any number of lattice sites.

\begin{acknowledgments}
D.F. would like to acknowledge financial support of the Federico Santa Maria Technical University
via the ``Programa 3: Apoyo a la Instalaci\'on en Investigaci\'on''.  
\end{acknowledgments}



\end{document}